\documentclass[twocolumn,showpacs,aps,prb,floatfix,nofootinbib,eqsecnum]{revtex4-1}
\usepackage{hyperref,bm}
\usepackage{amssymb,amsmath}
\usepackage{graphics}
\usepackage{graphicx}
\usepackage{subfigure}
\usepackage{tabularx}

\begin{document}

\newcommand{\contraction}[5][1ex]{%
	\mathchoice
		{\contractionX\displaystyle{#2}{#3}{#4}{#5}{#1}}%
		{\contractionX\textstyle{#2}{#3}{#4}{#5}{#1}}%
		{\contractionX\scriptstyle{#2}{#3}{#4}{#5}{#1}}%
		{\contractionX\scriptscriptstyle{#2}{#3}{#4}{#5}{#1}}}%
\newcommand{\contractionX}[6]{%
	\setbox0=\hbox{$#1#2$}%
	\setbox2=\hbox{$#1#3$}%
	\setbox4=\hbox{$#1#4$}%
	\setbox6=\hbox{$#1#5$}%
	\dimen0=\wd2%
	\advance\dimen0 by \wd6%
	\divide\dimen0 by 2%
	\advance\dimen0 by \wd4%
	\vbox{%
		\hbox to 0pt{%
			\kern \wd0%
			\kern 0.5\wd2%
			\contractionXX{\dimen0}{#6}%
			\hss}%
		\vskip 0.2ex%
		\vskip\ht2}}
\newcommand{\contracted}[5][1ex]{%
	\contraction[#1]{#2}{#3}{#4}{#5}\ensuremath{#2#3#4#5}}
\newcommand{\contractionXX}[3][0.06em]{%
	\hbox{%
		\vrule width #1 height 0pt depth #3%
		\vrule width #2 height 0pt depth #1%
		\vrule width #1 height 0pt depth #3%
		\relax}}

\newcommand{\im}{\text{i}}

\newcommand{\be}{\begin{equation}}
\newcommand{\ee}{\end{equation}}

\newcommand{\ev}[1]{\langle{#1}\rangle}
\newcommand{\vev}[1]{\langle 0|{#1}|0\rangle}

\newcommand{\up}{ \uparrow }
\newcommand{\dn}{ \downarrow }

\def\LSCO{La$_{2-x}$Sr$_x$CuO$_4$}
\def\LBCO{La$_{2-x}$Ba$_x$CuO$_4$}
\def\YBCO{YBa$_2$Cu$_3$O$_{6+x}$}
\def\HBCO{HgBa$_2$Cu$_3$O$_{4+\delta}$}
\def\BKBO{BaKBiO}
\def\C60{A$_x$C$_{60}$}
\def\LNSCO{La$_{1.6-x}$Nd$_{0.4}$Sr$_x$CuO$_{4}$}
\def\optimalLCO{La$_{1.85}$Sr$_{.15}$CuO$_4$}
\def\VO{V$_2$O$_3$}
\def\TMTSF{(TMTSF)$_2$X}
\def\ET{BEDT...}
\def\hty{high temperature superconductivity}
\def\hts{high temperature superconductors}
\def\ie{ {\it i.e.\/} }
\def\eg{ {\it e.g.\/} }
\def\sign{ {\rm sign } }
\def\SRO{ Sr$_{1+n}$Ru$_{n}$O$_{3n+1}$}
\def\SROone{ Sr$_{2}$Ru$_{}$O$_{4}$}
\def\SROtwo{ Sr$_{3}$Ru$_{2}$O$_{7}$}
\def\SROinf{ Sr$_{1}$Ru$_{}$O$_{3}$}
\def\CRO{Ca$_{1+n}$Ru$_{n}$O$_{3n+1}$}
\def\LCO{La$_2$CuO$_4$}
\def\LCOplus{La$_2$CuO$_{4+\delta}$}
\def\LSNiO{La$_{2-x}$Sr$_x$NiO$_{4+\delta}$}
\def\BSCCO{Bi$_2$Sr$_2$CaCu$_2$O$_{8+\delta}$}
\def\oxychloride{Ca$_{2-x}$Na$_x$CuO$_2$Cl$_2$}
\def\LNSCO{La$_{1.6-x}$Nd$_{0.4}$Sr$_x$CuO$_{4}$}
\def\TMTSFClO{(TMTSF)$_2$ClO$_4$}
\def\HgCu3{HgCa$_2$Cu$_3$O$_{8+y}$}
\def\HgCu4{HgBa$_2$Ca$_3$Cu$_4$O$_{10+y}$}
\def\TlCu{Tl$_2$Ba$_2$CuO$_{6+\delta}$}
\def\TlCu3{Tl$_2$Ba$_2$Ca$_2$Cu$_3$O$_{10+y}$}
\def\TlCu4{Tl$_2$Ba$_2$Ca$_3$Cu$_4$O$_{12+y}$}
\def\TlCun{Tl$_2$Ba$_2$Ca$_{n-1}$Cu$_n$O$_{2n+4+y}$}
\def\HgCun{HgBa$_2$Ca$_{n-1}$Cu$_n$O$_{2n+2+y}$}
\def\BiCun{Bi$_2$Sr$_2$Ca$_{n-1}$Cu$_n$O$_y$}
\def\BiCu3{Bi$_2$Sr$_2$Ca$_{2}$Cu$_3$O$_y$}
\def\BiCaMnO{Bi$_{1-x}$Ca$_x$MnO$_3$}
\def\NCCO{Ne$_{2-x}$Ce$_x$CuO$_{4\pm\delta}$}
\def\8LSCO{La$_{1.88}$Sr$_{.12}$CuO$_4$}
\def\110LNSCO{La$_{1.5}$Nd$_{0.4}$Sr$_{0.1}$CuO$_{4}$}
\def\stage4LCO{La$_{2}$CuO$_{4+\delta}$}
\def\Y248{YBa$_2$Cu$_4$O$_8$}
\def\PCCO{Pr$_{2-x}$Ce$_x$CuO$_{4\pm\delta}$}
\def\HTS{High temperature superconductors}
\def\htr{high temperature superconductor}
\def\cdw{charge-density wave}
\def\cdws{charge-density waves}
\def\Cdws{Charge-density waves}
\def\NbSe2{NbSe$_2$}
\def\TaSe2{TaSe$_2$}
\def\TiSe2{TiSe$_2$}
\def\NaCoOH2O{Na$_{0.3}$CoO$_{2y}$H$_2$O}
\def\MgB2{MgB${}_2$}
\def\telephone{Sr$_{14-x}$Ca$_{x}$Cu$_{24}$O$_{41}$}

\def\prl#1#2#3{Phys.\ Rev.\ Lett.\ {\bf #1}, #2 (#3)}
\def\pra#1#2#3{Phys.\ Rev.\ A {\bf #1}, #2 (#3)}
\def\prb#1#2#3{Phys.\ Rev.\ B {\bf #1}, #2 (#3)}
\def\prbrc#1#2#3{Phys.\ Rev.\ B {\bf #1} [RC], #2 (#3)}
\def\prd#1#2#3{Phys.\ Rev.\ D {\bf #1}, #2 (#3)}
\def\pre#1#2#3{Phys.\ Rev.\ E {\bf #1}, #2 (#3)}
\def\physrev#1#2#3{Phys. Rev. {\bf #1}, #2 (#3)}
\def\npb#1#2#3{Nucl.\ Phys.\ B {\bf #1}, #2 (#3)}
\def\npbfsold#1#2#3#4{Nucl.\ Phys.\ {\bf #1} [FS #2], #3, (#4)}
\def\npbfs#1#2#3{Nucl.\ Phys.\ {\bf #1} [FS], #2, (#3)}
\def\plb#1#2#3{Phys.\ Lett.\ B {\bf #1}, #2 (#3)}
\def\physrep#1#2#3{Phys.\ Rep.\ {\bf #1}, #2 (#3)}
\def\advphys#1#2#3{Adv.\ in Phys.\ {\bf #1}, #2 (#3)}
\def\mpla#1#2#3{Mod.\ Phys.\ Lett.\ A {\bf #1}, #2 (#3)}
\def\mplb#1#2#3{Mod.\ Phys.\ Lett.\ B {\bf #1}, #2 (#3)}
\def\ijmpa#1#2#3{Int.\ J.\ Mod.\ Phys.\ A {\bf #1}, #2 (#3)}
\def\ijmpb#1#2#3{Int.\ J.\ Mod.\ Phys.\ B {\bf #1}, #2 (#3)}
\def\rmp#1#2#3{Rev.\ Mod.\ Phys.\ {\bf #1}, #2 (#3)}
\def\jpc#1#2#3{J.\ Phys.\ C {\bf #1}, #2 (#3)}
\def\jpa#1#2#3{J.\ Phys.\ A {\bf #1}, #2 (#3)}
\def\physicac#1#2#3{Physica C {\bf #1}, #2 (#3)}
\def\physicaa#1#2#3{Physica A {\bf #1}, #2 (#3)}
\def\physicab#1#2#3{Physica B {\bf #1}, #2 (#3)}
\def\physicae#1#2#3{Physica E {\bf #1}, #2 (#3)}
\def\nature#1#2#3{Nature (London) {\bf #1}, #2 (#3)}
\def\natphys#1#2#3{Nature Physics {\bf #1}, #2 (#3)}
\def\natmat#1#2#3{Nature materials {\bf #1}, #2 (#3)}
\def\science#1#2#3{Science {\bf #1}, #2 (#3)}
\def\bams#1#2#3{Bull. Am. Math. Soc. {\bf #1}, #2 (#3)}
\def\baps#1#2#3{Bull. Am. Phys. Soc. {\bf #1}, #2 (#3)}
\def\cmp#1#2#3{Comm. Math. Phys. {\bf #1}, #2 (#3)}
\def\jmp#1#2#3{J. Math. Phys. {\bf #1}, #2 (#3)}
\def\jhep#1#2#3{J. High Ener. Phys. {\bf #1}, #2 (#3)}
\def\jstat#1#2#3{J. Stat. Mech.: Theor. Exp., {\bf #1}, #2 (#3)}
\def\jstatphys#1#2#3{J. Stat. Phys. {\bf #1}, #2 (#3)}
\def\annals#1#2#3{Ann. Phys (N.Y.) {\bf #1}, #2 (#3)}
\def\pnas#1#2#3{Proc. Natl. Acad. Sci. U.S.A. {\bf #1}, #2 (#3)}
\def\euro#1#2#3{Euro.\ Phys.\ Lett.\ {\bf #1}, #2 (#3)}
\def\europjb#1#2#3{Euro.\ Phys.\ Jour.\ B {\bf #1}, #2 (#3)}
\def\jpsj#1#2#3{J.\ Phys.\ Soc.\ Jpn.\ {\bf #1}, #1 (#3)}
\def\ssc#1#2#3{Solid State Comm.\ {\bf #1}, #2 (#3)}
\def\rpp#1#2#3{Rep. Prog. Phys.\ {\bf #1}, #2 (#3)}
\def\zetf#1#2#3{Zh. Eksp. Teor. Fiz.\ {\bf #1}, #2 (#3)}
\def\jetp#1#2#3{Sov. Phys. JETP.\ {\bf #1}, #2 (#3)}
\def\njp#1#2#3{New J. Phys.\ {\bf #1}, #2 (#3)}
\def\jphysa#1#2#3{J. Phys. A: Math. Theor. \textbf{#1}, #2 (#3)}
\def\jphyscmp#1#2#3{J. Phys.: Condens. Matter \textbf{#1}, #2 (#3)}

\title{Pair-Density-Wave Superconducting Order in  Two-Leg Ladders}
\author{Akbar Jaefari}
\author{Eduardo Fradkin}
\affiliation{Department of Physics, University of Illinois at Urbana-Champaign, 1110 West Green Street, Urbana, Illinois 61801-3080, USA}

\date{\today}
\begin{abstract}
We show using bosonization methods that extended Hubbard-Heisenberg models on two types of two leg ladders (without flux and  with flux $\pi$ per plaquette) 
have commensurate pair-density wave (PDW) phases. 
In the case of the conventional (flux-less) ladder the PDW arises when certain filling fractions for which commensurability conditions are met.
 For the flux $\pi$ ladder the PDW phase is generally present.
The PDW phase is characterized by a finite spin gap and a superconducting order parameter with a finite (commensurate in this case) wave vector and power-law 
superconducting correlations. In this phase the uniform superconducting order parameter, the $2k_F$ charge-density-wave (CDW) order parameter and the spin-density-
wave N\'eel order parameter exhibit short range (exponentially decaying) correlations. We discuss in detail the case in which the bonding band of the ladder 
 is half filled for which the PDW phase appears even at weak coupling. The PDW  phase is shown to be dual to a uniform superconducting (SC) phase with quasi long 
 range order. By making use of bosonization and the renormalization group we determine the phase diagram of the spin-gapped regime and study the quantum phase 
 transition. The phase boundary between PDW and the uniform SC ordered phases is found to be in the Ising universality class. We generalize the analysis to the case of 
 other commensurate fillings of the bonding band, where we find higher order commensurate PDW states for which we determine the form of the effective bosonized field 
 theory and discuss the phase diagram. We compare our results with recent findings in the Kondo-Heisenberg chain. We show that the formation of PDW order in the 
 ladder embodies the notion of intertwined orders.
\end{abstract}


\maketitle

\section{Introduction}

A spectacular dynamical layer decoupling of the transport properties accompanied by  different types of order parameters (charge, spin and superconducting) developing 
together has been observed experimentally in the stripe-ordered (or nearly ordered)  cuprate superconductor {\LBCO} 
at zero external magnetic field\cite{li-2007,tranquada-2008}, in underdoped {\LSCO} at moderate magnetic fields, and in optimally-doped {\LBCO} 
at low magnetic fields.\cite{wen-2011} 
A sequence of phase transitions are seen in these materials with the ``normal'' to charge-stripe ordered transition occurring first, followed by a spin-stripe order
 transition with a lower critical temperature. 
 For instance, in {\LBCO} near doping $x=1/8$  a spectacular decoupling of the layers develops in transport measurements with the ratio of the c-axis  
 $\rho_c$ to the ab-plane $\rho_{ab}$ resistivities, becoming larger than $10^5$, begins to develop quite rapidly at temperatures right below the 
 spin-ordering transition. At a critical temperature of the order of $T_c^{2D}\sim 20 K$ (depending on the precise doping) the copper oxide planes appear 
 to become superconducting while the c-axis transport remains resistive. The full three-dimensional resistive transition is seen only below $10 K$. 
 A superconducting state with a Meissner effect and (presumably) $d$-wave superconductivity is seen below  $T_c^{3D}\sim 4K$. However, even though t
 he critical temperature of the uniform $d$-wave superconducting state is much lower near $x=1/8$ than for other doping levels 
 (where it is typically $\sim 40 K$), the experiments show that the anti-nodal superconducting gap is essentially unsuppressed.\cite{valla-2006,he-2008}

A strikingly similar transport anisotropy has been observed very recently in the temperature-pressure phase diagram of the heavy fermion superconductor 
CeRhIn$_5$.\cite{park-2011}  In this strongly correlated material the orders that develop are conventionally identified as a spin-density wave metallic state 
and a uniform $d$-wave superconductor. In the phase in which both orders coexist the ratio $\rho_c/\rho_{ab}$ becomes large ($ \sim 10^3$) with $\rho_{ab}$ 
eventually becoming unmeasurably small as is the superconductivity became two-dimensional (as in the  case of {\LBCO} near $1/8$ doping\cite{li-2007}). 

The most unusual aspect of these experiments is not just the existence of multiple coexisting orders, but the dynamical layer decoupling seen in transport. 
That is, the existence of a significant temperature range over which there is a form of two-dimensional superconductivity in the planes but which are otherwise 
decoupled as if there was no Josephson effect between them. 

Berg {\it et al}\cite{berg-2007,berg-2009a,berg-2009b,berg-2009c} showed that these seemingly contradictory results can be explained if one assumes that in 
this state  charge, spin and superconducting orders are not competing with each other but rather that they are intertwined, with the superconducting state in 
the planes  also being striped, {\it i.e.} it is a unidirectional {\em pair density wave} (PDW) with the property that the phase of the superconducting order 
parameter alternates in sign, as if the axes of the $d$-wave order parameter were to rotate by $90^\circ$ (with vanishing average value for the superconducting 
order parameter). 
The order parameter for the PDW  state  is
\be
	\Delta_\text{PDW}(\vec x) = \Delta_{\vec{Q}_\text{PDW}}e^{\im \vec Q_\text{PDW} \cdot\vec x} +\Delta_{-\vec{Q}_\text{PDW}}e^{-\im \vec Q_\text{PDW}\cdot\vec x}
	\label{eq:PDW}
\ee
with the PDW ordering wave vector $\vec Q_\text{PDW}$ pointing in the direction normal to the stripes, and (in the LTT lattice structure of {\LBCO}) rotates by 
$90^\circ$ from plane to plane. Translation invariance of the underlying system further dictates that the  ordering wave vectors for spin 
($\vec Q_\text{SDW}$), charge ($\vec Q_\text{CDW}$) and superconducting ($\vec Q_\text{PDW}$)  order parameters obey the relation 
$2\vec Q_\text{PDW}=2\vec Q_\text{SDW}=\vec Q_\text{CDW}$. 

In this paper we show that PDW type states do occur in the phase diagram of strongly correlated systems on two-leg ladders. 
We focus on these systems since in this context the physics of strong correlations can be controlled and there are powerful bosonization and numerical 
methods to investigate their phase diagrams and to compute their correlators. For reasons that will be explained below, states of this type generally 
involve strong correlation physics which is much harder to control in two dimensions. Results obtained  in one-dimensional systems, but with lots of caveats, 
can be used to develop at least qualitatively theories of two-dimensionally ordered states. In this work we will use bosonization methods, which are accurate 
at relatively weak coupling, to show how intertwined orders of this type develop in two-leg ladder. This work in many ways is an extension of the results of 
Ref.[\onlinecite{berg-2010}] that showed that  the PDW state represents the spin-gapped phase (``Kondo-singlet'')  of the Kondo-Heisenberg chain. 

Non-uniform superconducting states were proposed by Fulde and Ferrell\cite{fulde-1964} and by Larkin and Ovchinnikov\cite{larkin-1964}, and are conventionally 
called FFLO states. The superconducting component of the PDW state discussed above has the form of a Larkin-Ovchinnikov state.\cite{larkin-1964}  The   
Fulde-Ferrell state\cite{fulde-1964}  has spiral order. 
FFLO states have been proposed over the years for different types of superconducting materials, most recently in the phase diagram  of the heavy fermion 
superconductor CeCoIn$_5$ at finite magnetic fields,\cite{kenzelmann-2008} although the experimental evidence for them is weak (at best). 
There are also recent theoretical proposals for FFLO-type states in cold atomic fermionic systems with unbalanced populations,\cite{radzihovsky-2009} 
and in quantum wires of multi-valley semiconductors.\cite{datta-2009} 

In the conventional  theory of FFLO states one assumes a BCS-type system (a Fermi liquid) in which the spin up and down Fermi surfaces are split by the 
Zeeman interaction with an external magnetic field. As a result the ordering wave vector of the FFLO states is tuned by the Zeeman interaction and, 
hence by the magnitude of the magnetic field. A problem with this mechanism is that  the usual nesting of the Fermi surface that leads to a BCS state 
zero-momentum pairing is generally lost if the Fermi surfaces are split,  and the superconducting states with finite-momentum pairing can only happen 
for large enough attractive interactions, instead of being an infinitesimal instability as in the conventional BCS case.\cite{schrieffer-1964} FFLO phases driven by the Zeeman interaction (and hence with a spin- imbalance and a broken $SU(2)$ spin symmetry) in quasi-one-dimensional systems, including ladders, were discussed in Ref. [\onlinecite{Roux-2007b}].

On the other hand the PDW state is the result of strong correlation physics, does not involve a Zeeman interaction, and does not occur at weak coupling.  
In the perspective of Ref.[\onlinecite{berg-2009a}], the PDW state is another manifestation of the concept of frustrated phase separation which is behind the 
development of inhomogeneous electronic states in strongly correlated materials,\cite{emery-1993} whose result are electronic liquid crystal 
phases\cite{kivelson-1998} of which the PDW is a particularly interesting example. An understanding of the physics of this state should shed 
light on the connection between electronic inhomogeneity and superconductivity.

A BCS-mean-field-theory of a extended Hubbard  model in two dimensions has been developed by Loder and coworkers\cite{loder-2009,loder-2011} 
who  found that, as part of a rich phase diagram, the PDW  is the ground state for large enough interactions. Earlier mean field theory results  in the $t-J$ model 
 by Yang {\it et al}\cite{yang-2009} found that the PDW is energetically  very close to being the ground state. These authors found that (within their mean field theory) 
 the ground state is a modulated $d$-wave superconductor, {\it i.e.} a state in which the uniform $d$-wave order parameter coexists with the PDW state, 
 {\it i.e.} a state in which superconductivity and stripe order coexist.\cite{Jaefari-2010} 
 Several variational Monte Carlo calculations in the 2D $t-J$ model have found that this is a very competitive state near doping $x=1/8$ although 
 not quite the best variational ground state.\cite{himeda-2002,raczkowski-2007,capello-2008}. 
While these results are encouraging they suffer from the problem that either they use approximations that are not controlled  in these strongly coupled regimes 
(as in the mean field studies), or that the search yields the best variational state within a restricted class which would be adequate at weak coupling 
but not in these regimes. Nevertheless these results indicate that from the point of view of local energetics the PDW state is very competitive and 
would most likely be the ground state for some extension of the Hamiltonians that were studied. 
A very recent and numerically intensive calculation using infinite projected entangled pair states (an extension of the density matrix renormalization group) 
by Corboz {\it et al}\cite{corboz-2011} have found stripe states coexisting with superconductivity in the 2D $t-J$ 
model but has not yet investigated the presence (or absence) of a PDW state.

In this paper we revisit the problem of the phase diagram of extended Hubbard-Heisenberg type models on two-leg ladders using bosonization methods. Here we show 
that  the two-leg ladder has a phase that can be identified with a version of the PDW state, another phase in which uniform superconducting order is present and there is a 
continuous phase transition between these two phases. The two-leg ladder is an ideal model-system to study since it is well known from DMRG and bosonization results 
to have broad regimes of coupling constants and doping in which the ground state has a spin-gap and $d$-wave superconducting correlations, as well as a strong 
tendency to stripe order.\cite{noack-1994,noack-1997,white-1998a,white-2000} Using a weak coupling band terminology, the PDW state is stabilized when one of the 
band of the ladder, say the bonding band, is at special commensurate fillings.

While there are extensive bosonization studies of the phase diagram of two-leg ladders,\cite{wu-2003,giamarchi-2003} the existence of phases with 
PDW order has not been previously investigated. 
Here we show that there is a connection between two leg ladder and the Kondo-Heisenberg (KH) chain, a model  of a 1D electron system  interacting 
with a spin-$1/2$ Heisenberg antiferromagnetic chain through Kondo interaction.  
In a series of papers, Zachar et.~al.~\cite{zachar-1996, zachar-2001, zachar-2001b} showed the existence of a phase in which the correlations of the uniform 
superconducting order parameter decays exponentially fast while the correlations of a staggered superconducting state (composite SC) has quasi long range order. 
Berg {\emph {et.~al.}}~\cite{berg-2010} studied the KH problem using the Density Matrix Renormalization Group (DMRG) calculations and showed that  the spin-gapped 
phase of the KH chain \cite{sikkema-1997} is a commensurate PDW state with $\Delta_{\vec{Q}} = \Delta^*_{-\vec Q}$. 
Similarly, the PDW phases that we find in the Hubbard-Heisenberg model are also commensurate and have wave vector $Q_{PDW}=\pi$. In addition, and similarly to 
what happens in the KH chain, the PDW order parameter is a composite operator of a triplet uniform SC order parameter of the anti-bonding band and the 
antiferromagnetic (N\'eel) order parameter of the bonding band. Separately, these two order  parameters have short range order in the PDW state.
In the PDW phase in the ladder system translation symmetry is broken spontaneously whereas in the KH chain it is broken explicitly by the spacing 
between the static spins. We discuss in detail the 
quantum phase transition between the PDW state and the phase with uniform SC order and it is found to be in the  universality class of the quantum Ising chain.

We also consider an extended Hubbard-Heisenberg model on a two-leg ladder with flux $\Phi$ per plaquette. An important difference of this ladder system is that for 
general flux time reversal symmetry is broken explicitly except for the spacial case of flux $\Phi=\pi$ per plaquette which is time reversal invariant. The presence of a non-
zero flux changes the band structure by doubling the number of Fermi points at which the bonding band crosses the Fermi energy. We have not explored in full the 
complex phase diagram of this system except for the case of flux $\Phi=\pi$. Here we found that for generic fillings of the bonding band the system obeys an Umklapp 
condition which leads to a spin gap state. We explored the phase diagram in this case and found that it generally supports two types of uniform and PDW 
superconducting orders. In this system the PDW order parameters are bilinears in fermion operators and are not composite operators and in the previous case. Here, as in 
the conventional ladder and in the KH chain, there is no coexistence between PDW and uniform orders: when one order develops (which in 1D means power law 
correlations) the correlators of the other order parameters decay exponentially at long distances. In addition to the SC phase we also found four incommensurate CDW 
phases. The quantum critical behavior of this system is more complex, reflecting the larger diversity of phases that we encountered. In particular while the generic 
quantum phase transitions are also in the universality class of the quantum ising chain, for some special choices of parameters the symmetry associated with the quantum 
critical behavior is enlarged to $U(1)$ and it is now in the universality class of a spinless Luttinger model. 

Two-leg ladders with flux $\Phi$ per plaquette were studied (both analytically and using numerical DMRG methods) by Roux and coworkers\cite{Roux-2007} in their work 
on diamagnetic effects in two-leg ladders, as well as by Carr and coworkers\cite{Narozhny-2005,Carr-2006} who used bosonization methods to study many aspects of the 
phase diagram.  However in their work these authors did not consider the case of flux $\Phi=\pi$ per plaquette in which time reversal invariance plays a key role, and  the 
problem of PDW phases, that (as we show occur here) generically at flux $\Phi=\pi$, and  which is the focus of the present work. 

This paper is organized as follows. In section \ref{sec:model} we introduce the model of the two-leg ladder and its effective field theory using bosonization methods.
We can draw the phase diagram using the microscopic parameters of the ladder in the the weak coupling regime where their relation with  the coupling constants of the 
bosonized theory is known explicitly. Although the form of the effective field theory does not change with the strength of the coupling constants, in more strongly coupled 
regimes numerical methods must be used to established this relation.
In section \ref{sec:half-filled} we present the bosonized theory of a ladder whose bonding band is half-filled. Here we show that this effective low energy theory has a 
hidden self-duality.
In section \ref{sec:phase-diagram-half} we use renormalization group methods  to determine the phase diagram for the  case of a half-filled bonding band, and show that, 
in addition of a Luttinger Liquid type phase, it also has two SC states, one with uniform SC order and the other with PDW order with wave vector $Q_\text{PDW}=\pi$. We 
show that in this case there is a direct quantum phase transition between the phase with uniform SC order and the PDW phase which is in the universality class of the 
quantum Ising chain.. 
 In section \ref{sec:other-commensurabilities} we extend this analysis to regimes with a bonding band at other commensurate fillings. The resulting  PDW phase with 
 wave vector $Q_\text{PDW}$ coexists (or is intertwined) with an also commensurate charge-density-wave (CDW) state in the bonding band with wave vector 
 $Q_\text{CDW}=Q_\text{PDW}/2$. Unlike the half-filled case, this state does not occur at weak coupling. In section \ref{sec:flux} we consider an extended 
 Hubbard-Heisenberg model on a  two-leg ladder with flux $\Phi$ per plaquette. Here we show that the commensurate PDW phase arises naturally in this 
 frustrated band structure although through a different  mechanism. The conclusions are presented in section \ref{sec:conclusions}. 
 The RG equations for the general case are presented in Appendix \ref{sec:RG-pi-flux}.
 The solution of the effective field theory of the flux $\Phi=\pi$ model for special combinations of coupling constants and refermionization is given in 
 Appendix \ref{sec:refermionization-pi-flux}.

\section{Model of the Two-Leg Ladder and Effective Field Theory}
\label{sec:model}

Consider a model of the two-leg ladder whose Hamiltonian is $H=H_0+H_\text{int}$. The kinetic energy term is
\begin{align}
	H_0=& -t\sum_{i,j,\sigma}\left\{ c^\dagger_{i,j, \sigma}c_{i,j+1,\sigma} + \text{h.c.} \right\} \nonumber\\
	&-t_\perp \sum_{j,\sigma}\left\{ c^\dagger_{1,j, \sigma}c_{2,j,\sigma} + \text{h.c.} \right\}
\end{align}
with $t$ and $t_\perp$ being, respectively, the intra-leg and inter-leg hopping amplitudes, and $i=1,2$ being the chain index and $j$ being the lattice site index .
The interaction terms of the ladder Hamiltonian have the form of a extended 
Hubbard-Heisenberg model,
\begin{align}
\begin{split}
	H_\text{int}= &~ U\sum_{i,j} n_{i,j,\up}n_{i,j,\dn} + V_\parallel \sum_{i,j} n_{i,j}n_{i,j+1}\\
	 + V_\perp & \sum_{j} n_{1,j}n_{2,j} +V_d  \sum_{j} \left\{ n_{1,j}n_{2,j+1}+n_{1,j+1}n_{2,j} \right\}\\
		& + J_\parallel \sum_{i,j} \vec S_{i,j} \cdot \vec S_{i,j+1} + J_\perp \sum_{j} \vec S_{1,j} \cdot \vec S_{2,j} \\
		&+ J_d  \sum_{j} \left\{ \vec S_{1,j} \cdot \vec S_{2,j+1}+\vec S_{1,j+1} \vec S_{2,j} \right\}
		\label{lattice-model}
\end{split}
\end{align}
where $U$ is the on-site Hubbard repulsion, $V_\parallel$, $V_\perp$ and $V_d$ are the  nearest neighbor and next-nearest neighbor ``Coulomb'' repulsions, 
and $J_\parallel$, $J_\perp$ and $J_d$ are the nearest and next-nearest neighbor exchange interactions.

In  the weak coupling regime, $U,V,J \ll t,t_\perp$,  we proceed by  first  diagonalizing the kinetic energy term $H_0$ and finding its low-energy spectrum. 
This can be done by switching to the bonding and anti-bonding basis defined as  (at each rung $j$ of the ladder and for each spin polarization $\sigma$)
\begin{equation}
c_{b,a}=\frac{1}{\sqrt{2}}(c_{2}\pm c_{1})
\label{eq:bonding-antibonding}
\end{equation}
 In the new basis the kinetic term reads as
\begin{align}
	H_0 = \sum_{\eta=a,b} \sum_{j,\sigma}t_\eta \left\{ c^\dagger_{\eta,j,\sigma} c^{}_{\eta,j+1,\sigma} + \text{h.c.}\right\}
\end{align}
in which $b$ and $a$ stand for bonding and anti-bonding, and where $t_\eta=t \pm t_\perp$ for $\eta=b,a$ respectively. 

\begin{figure}[hbt]
		\includegraphics[width=0.4\textwidth]{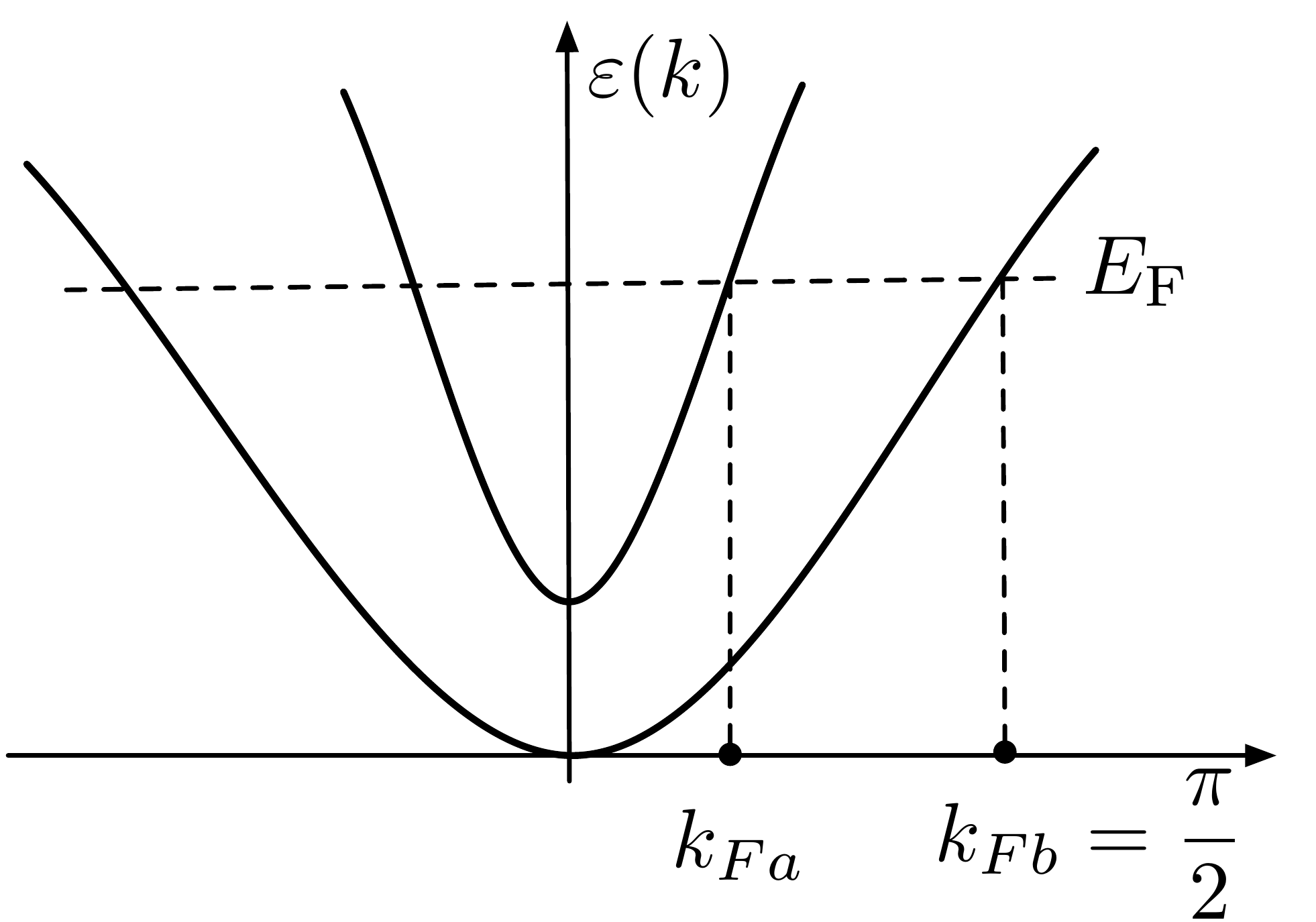}
		\caption{Schematic picture of the bonding and anti-bonding bands. The  bonding band $b$ is kept at half-filling. 
		The filling of the anti-bonding band $a$ is general. Here $k_{F,a}$ and $k_{F,b}=\frac{\pi}{2}$ are the Fermi points for each band.}
		\label{fig: bonding-antibonding}
\end{figure}


In order to find the continuum limit representing 
the low-energy and long-wavelength behavior of the model, we linearize the energy dispersion of each band of the ladder around the respective Fermi wave vector 
\begin{equation}
\varepsilon_\eta(k) \approx E_F+v_\eta (k-k_{F,\eta}) 
\label{eq:dispersion-modes}
\end{equation}
where $v_\eta=2t_\eta\sin(k_{F,\eta})$ are the Fermi velocities, $k_{F,\eta}$ the Fermi wave vectors for each band, and $E_F$ is the Fermi energy.
We now consider  the regime of  small fluctuations close to the  Fermi points of each band:
\be
	\frac{1}{\sqrt{a}}c_{\eta,j,\sigma} \rightarrow R_{\eta,\sigma}(x)e^{\im k_{F\eta}x} + L_{\eta,\sigma}(x)e^{-\im k_{F\eta}x},
\ee
where $R$ and $L$ are right- and left-moving components of the electron field, $x=ja$ is the position, and $a$ is the lattice constant (the rung spacing). In this limit, the 
kinetic term of the Hamiltonian takes the standard continuum form 
\be
	H_0 = \sum_{\eta,\sigma} \int dx (-\im v_\eta)\left\{ R^\dagger_{\eta,\sigma} \partial_x R_{\eta,\sigma} - L^{\dagger}_{\eta,\sigma} \partial_x L^{}_{\eta,\sigma}  \right\}.
\ee

The most general continuum interacting Hamiltonian density (up to possible Umklapp processes that will be discussed below) compatible with the charge conservation 
and the global $SU(2)$ spin symmetry with scaling dimension two (marginal) operators  has the following form
 \begin{align}
 \begin{split}
  	{\cal H}_\text{int} =& \sum_{\eta=a,b} \left\{ f_{c1\eta}  \left(J^2_{R\eta}+J^2_{L\eta}\right) + g_{c1\eta} \, J_{R\eta}J_{L\eta} \right\}\\
		&+ \sum_{\eta=a,b} \left\{ f_{s1\eta} (\vec J^{~2}_{R\eta}+ {\vec J}^{~2}_{L\eta}) + g_{s\eta} \, \vec J_{R\eta} \cdot \vec J_{L\eta} \right\}\\
		&+ f_{c2} \left(J_{Ra}J_{Rb}+J_{La}J_{Lb}\right)\\
		&+ g_{c2} (J_{Ra}J_{Lb}+J_{La}J_{Rb}) \\
		&+ f_{s2}(\vec J_{Ra}\cdot \vec J_{Rb} + \vec J_{La}\cdot \vec J_{Lb})\\
		&+ g_{s2} (\vec J_{Ra}\cdot \vec J_{Lb}+\vec J_{La}\cdot \vec J_{Rb} )\\
		&+ \lambda_t ( \vec \Delta^\dagger_b \cdot \vec \Delta_a + \text{h.c.} ) + \lambda_s( \Delta^\dagger_b\Delta_a + \text{h.c.})\label{smooth-continuum-H}
\end{split}
\end{align}
in which $J_{R/L,\eta}$ are the right and left moving components of the charge density $J_\eta=J_{R,\eta}+J_{L,\eta}$ for each band $\eta$
\be
 	J_{R,\eta}= \sum_{\sigma} R^\dagger_{\sigma,\eta}  R^{}_{\sigma,\eta}, \quad J_{L,\eta} = \sum_{\sigma} L^\dagger_{\sigma,\eta} L^{}_{\sigma,\eta}
\ee
 $\vec J_{R/L,\eta}$ are right and left moving components of spin density $\vec J_\eta = \vec J_{R,\eta} + \vec J_{L,\eta}$ for each band $\eta$,
\be
 	\vec J_{R,\eta}= \frac{1}{2} \sum_{\alpha\beta} R^\dagger_{\alpha,\eta} \vec \sigma_{\alpha\beta}R^{}_{\beta,\eta}, 
	\quad \vec J_{L,\eta} = \frac{1}{2}\sum_{\alpha\beta} L^\dagger_{\alpha,\eta} \vec\sigma_{\alpha\beta}L^{}_{\beta,\eta}.
\ee
where the components of  the vector $\vec \sigma$ are the three Pauli spin matrices, and 
\begin{equation}
\Delta_\eta = L_{\up,\eta} R_{\dn,\eta} - L_{\dn,\eta} R_{\up,\eta}, \qquad \vec \Delta_\eta =
\sum_{\alpha,\beta} L_{\alpha,\eta}(\im \vec\sigma \sigma_y)_{\alpha\beta} R_{\beta,\eta}
\end{equation}
are singlet and triplet pairing operators respectively for each band $\eta$. 

In the weak coupling limit, the relation between the coupling constants of the continuum theory of Eq.\eqref{smooth-continuum-H} and the parameters of the Hamiltonian 
of the microscopic lattice model of Eq.\eqref{lattice-model} can be found through this naive continuum limit procedure. 
This has been done before and can be found, for example, in  Ref.~[\onlinecite{wu-2003}]. 
We note that here by keeping only (naively) marginal operators we have neglected a host of irrelevant operators that are present in the lattice model 
that do not change the form of the low energy theory (although change the definition of the coupling constants by small amounts).
In the intermediate to strong coupling limit, either non-perturbative methods such as Bethe ansatz (applicable only if the system is integrable) or numerical 
density-matrix renormalization group (DMRG) calculations are required to make a quantitative connection between the the lattice model and the effective  
continuum field  theory that we will use below. Nevertheless the form of ${\cal H}_\text{int}$ is general as seen here. 

Here we will be interested in the case of a half-filled  bonding band.  This situation will happen naturally as the total filling of the ladder is varied without 
breaking any symmetries of the ladder. we should note that if one wanted to specify the filling of the bonding and anti-bonding bands separately,
 it would be necessary to set a chemical potential difference between the two legs which would make the ladder asymmetric.  
 At any rate, for a half-filled bonding band, in addition to the interactions presented in ${\cal H}_\text{int}$, we need to include the following (bonding band) 
 Umklapp process
\be
	{\cal H}_{u,b} = g_{u,b} ( L^\dagger_{b\up}L^\dagger_{b\dn}R^{}_{b\dn}R^{}_{b\up}e^{\im4k_{Fb}x} + \text{h.c.})
\ee
The value of  $g_{u,b}$ in the weak coupling regime is found to be given by
\be
	\frac{1}{a} g_{u,b} = \frac{1}{2}(U+V_\perp) - (V_\parallel+V_d) -\frac{3}{8} J_\perp+ \frac{3}{4} (J_\parallel+J_d)
\ee
We will furthermore assume that for a substantial range of parameters of interest ${\cal H}_{u}$ is marginally relevant for a half filled bonding band. 
We will get back to this point in the next section.

\section{Analysis of a two-leg ladder system with a half-filled bonding band}
\label{sec:half-filled}

We will now consider in detail the case when the bonding band is half filled and its Fermi wave vector is $k_{Fb}=\pi/2$. 
In this case there is a marginally relevant interaction representing the Umklapp process mentioned above. The main effect of this process is to open a  
charge gap $\Delta_c$ in the bonding band. 
Therefore for the energies much smaller than the charge gap, one can assume that the charge degrees of freedom on bonding-band $b$ are frozen-out and 
hence play no roll in the low energy limit of the remaining degrees of freedom. Moreover, due to the charge gap in the bonding band, 
all the interactions with net charge transfer between the bands, namely singlet and triplet SC in Eq. \eqref{smooth-continuum-H} processes, become irrelevant. 
Therefore the only remaining charge degree of freedom, which is that of the anti-bonding band $a$, is decoupled from the rest of the dynamics, 
and this being a one-dimensional system, the effective field theory of the charge sector of the anti-bonding band is described by the Luttinger Liquid (LL) theory. 
In its bosonized form the effective  Hamiltonian density for the charge sector involves the Bose field $\phi_c$ and its dual field $\theta_c$ for the anti-bonding 
band $a$ only (where to simplify the notation we dropped the label) and reads reads as
\be
	{\cal H}_c = \frac{v_c}{2} \left\{ K_c(\partial_x\theta_c)^2+\frac{1}{K_c}(\partial_x\phi_c)^2 \right\}
	\label{eq:Hc}
\ee
where $K_c$ is the charge Luttinger parameter of the system and $v_c$ is the velocity. The fields $\phi_c$ and $\theta_c$ are dual to each other and satisfy the standard 
equal-time canonical commutation relations
\begin{equation}
[\phi_c(x),\partial_x\theta_c(x')]=\im\delta(x-x')
\end{equation}
which identifies $\Pi_c=\partial_x \theta_c$ with the canonical momentum.

 All the possible remaining interactions between the degrees of freedom of both bands are in the spin sector. The most general interacting Hamiltonian for the spin sector, 
 which is symmetric under the exchange of the band index $\eta=a,b$, is
\begin{align}
  	{\cal H}_s &= - g_{s1}(\vec J_{Rb} \cdot \vec J_{Lb}+\vec J_{Ra} \cdot \vec J_{La})\\
		&\qquad - g_{s2} (\vec J_{Rb}\cdot \vec J_{La}+\vec J_{Lb}\cdot \vec J_{Ra})
\end{align}
Following the standard bosonization procedure in one dimension, in terms of spin boson fields 
\begin{equation}
\phi_{s\pm} = \frac{1}{\sqrt{2}}(\phi_{s,b} \pm \phi_{s,a})
\end{equation} 
we arrive at the following form for ${\cal H}_s$
\begin{align}
 	&{\cal H}_{s} = \frac{v_{s\pm}}{2} \left[ K_{s\pm}(\partial_x \theta_{s\pm})^2 + K^{-1}_{s\pm}(\partial_x \phi_{s\pm})^2 \right] \label{eq:shamiltonian}\\
		&+\frac{\cos(\sqrt{4\pi}\phi_{s+})}{2(\pi a)^2} \left[ g_{s1} \cos(\sqrt{4\pi}\phi_{s-}) + g_{s2} \cos(\sqrt{4\pi}\theta_{s-})\right] \nonumber
\end{align}
where the Luttinger parameters $K_{s\pm}$ and velocities $v_{s\pm}$ are related to $g_{s\pm} = (g_{s,1}\pm g_{s,2})/2$ as
\be
	K_{s\pm} = \sqrt{\frac{2\pi v_f+ g_{s\pm}}{2\pi v_f- g_{s\pm}}}, \qquad v_{s\pm} = \sqrt{v^2_f-\left(\frac{g_{s\pm}}{2\pi}\right)^2}.
	\label{eq:Ks-}
\ee
in which $v_f$ is the Fermi velocity of the noninteracting problem. The dual fields $(\theta_{s\pm},\phi_{s\pm})$ obey similar commutation relations as the dual fields in the 
charge sector. In general $g_{s1}$ is different for each band 
$g_{1b}\neq g_{1a}$. This introduces terms involving operators of the form $ \partial_x \phi_{s+}\partial_x \phi_{s-}$ and $ \partial_x \theta_{s+} \partial_x \theta_{s-}$. 
Although these 
are marginal operators we will neglect them for now since we will later argue that the results are not essentially affected by these terms in phases with a spin gap.  
In the absence of the spin gap, {\it i.e.} in the Luttinger Liquid phase, these operators change (among other things) the scaling dimensions of the 
observables.\cite{emery-2000}

Upon inspecting the Hamiltonian of the spin sector Eq. \eqref{eq:shamiltonian}, we see that it is invariant under the duality transformation 
\begin{equation}
(\phi_{s-},\theta_{s-}, g_{s1},g_{s2},K_{s-})\rightarrow(\theta_{s-},-\phi_{s-},g_{s2},g_{s1},K^{-1}_{s-})
\label{eq:duality}
\end{equation}
 Thus, this duality symmetry guarantees the existence of a dual phase associated with
the vanishing of the coupling constant $g_{s2}=0$. We will see that in contrast to the PDW phase which is controlled by KH fixed point, in the dual phase 
the uniform SC is 
the most dominant instability. We will discuss the implications of this symmetry on the phase diagram later on.

In the $g_{s1}=0$ limit  the Hamiltonian of Eq.\eqref{eq:shamiltonian} turns out to be  the same as the effective field theory description of the continuum 
limit of the one-dimensional 
Kondo-Heisenberg (KH) chain\cite{zachar-2001,zachar-2001b,berg-2010} with nearest-neighbor Kondo spins. The KH chain is a system of a 1DEG 
(which usually is taken to be non-interacting but this is not necessary) and a one-dimensional array of spin-$1/2$ degrees of freedom, 
a 1D quantum antiferromagnet with exchange interaction $J_H$. The spacing between the spin degrees of freedom defines the unit cell of the chain and in 
general it is not equal to the lattice spacing of the 1DEG. The coupling between the spin chain and the 1DEG is the Kondo exchange coupling $J_K$. 

That  these two problems have almost the same low energy effective field theory of the same form is not a coincidence since there is 
a formal analogy between the two problems. In both 
cases we have a gapless 1DEG, the free fermion band of the KH problem with the 1DEG of the electrons in the anti-bonding $a$-band of the two leg ladder system, which 
in both cases is coupled to a Heisenberg spin-$1/2$ chain. It is known that the KH chain (regardless of the spacing between the Kondo spins) 
has a broad regime of its phase diagram in which there is a spin gap.\cite{sikkema-1997} This phase has been identified with a commensurate 
PDW phase.\cite{berg-2010} One difference between these two systems  is that in the two-leg ladder the coupling is the tunneling matrix element $t_\perp$ whereas 
in the KH case it is the local Kondo exchange $J_K$. However since the bonding band of the ladder has a charge gap the only active coupling allowed at low energies is 
also the effective exchange coupling. Thus in the $g_{s1}=0$ limit both problems are the same.  We will see below that in the regime with $g_{s1}=0$ the 
parameters of the Hamiltonian flow under the renormalization group  to a stable fixed point of the characterized by pair density wave correlations.
Also, in the  $g_{s1}=0$ limit the system has an exponentially small gap (which can be determined form a mapping to the $SU(2)$ Thirring model~\cite{zachar-2001})
that is stable against small perturbations of the form we discussed  above~\cite{emery-1979,gogolin-1998}. 

However,  there is an important qualitative difference between these two systems. 
While the Kondo-Heisenberg chain is translationally invariant only if the lattice spacing of the quantum antiferromagnet (the distance between the Kondo spins) 
is the same as the lattice spacing of the 1DEG, whereas the ladder is a translationally invariant system in all cases. This will play an important role in our analysis.


\section{Phase Diagram of the system with a half-filled bonding band}
\label{sec:phase-diagram-half}

\begin{figure}[hbt]
		\includegraphics[width=0.45\textwidth]{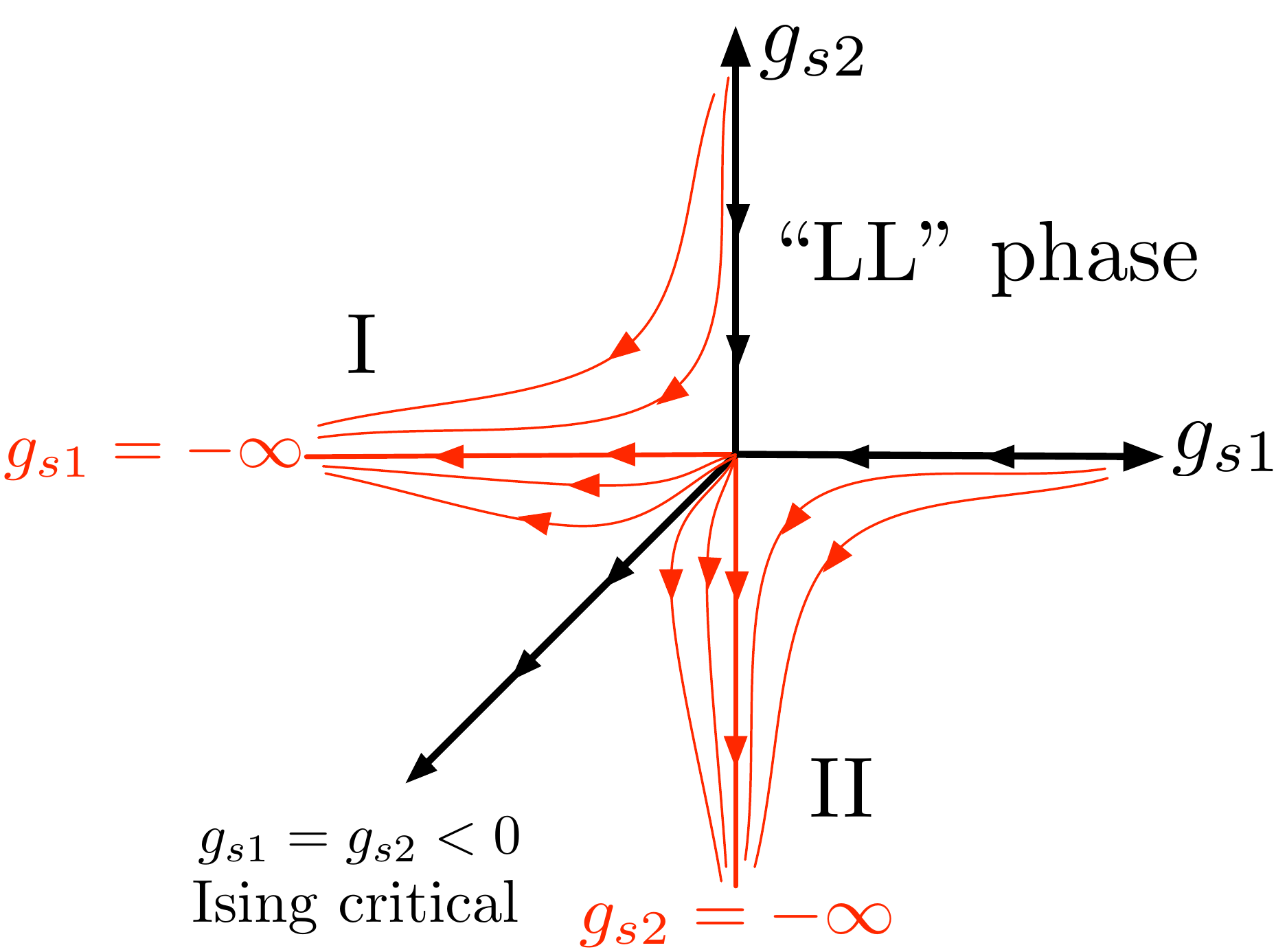}
		\caption{Schematic phase diagram when the bonding band of the ladder is half-filled, 
		shown as a projection of the $SU(2)$-invariant RG flows onto the $(g_{s1},g_{s2})$ plane. 
		The solid black lines represent the separatrix between different phases. 
		Due to the duality symmetry  in the effective Hamiltonian for the spin sector, the phase diagram  is symmetric around $g_{s-}=0$. 
		The quadrant $g_{s1},g_{s2} >0 $ flows into the Gaussian fixed point $g_{s1}=g_{s2}=0$ and is a Luttinger Liquid (LL) phase. 
		Region II is controlled by the PDW strong coupling fixed point at $(g_{s1}=0,g_{s2}=-\infty)$.
		Region I is controlled by decoupled spin-gapped fixed point of $(g_{s1}=0,g_{s2}=-\infty)$. 
		There is a KT transition from the LL behavior across the half-line $g_{s2} =0$ and $g_{s,1}>0$ to region II, and across the half-line 
		$g_{s1} =0$ and $g_{s,2}>0$ to region I. 
		The half-line $g_{s1}=g_{s2}<0$ represents a quantum phase transition between the two strong-coupling phases and it is in the Ising universality class.}
\label{fig:SU(2)-RG}
\end{figure}

We will now discuss in detail the phase diagram and phase transitions of a extended Hubbard-Heisenberg model on a ladder in which one band, 
the bonding $b$ band is half filled, as shown schematically in Fig.\ref{fig:SU(2)-RG}.

\subsection{Weak Coupling RG analysis}
\label{sec:RG}

The total effective  Hamiltonian is ${\cal H}={\cal H}_c+{\cal H}_s$, where the Hamiltonian density for the charge sector ${\cal H}_c$ is given by Eq.\eqref{eq:Hc} and the 
Hamiltonian density of the spin sector ${\cal H}_s$ is given by Eq.\eqref{eq:shamiltonian}. The total Hamiltonian has  five coupling/parameters, $K_c$, $K_{s\pm}$, and 
$g_\pm$ (or equivalently $g_{s1}$ and $g_{s2}$). The Luttinger parameter of the charge sector will not renormalize since the charge sector decouples, while all the 
parameters in the spin sector are subject to renormalization. The  one-loop RG equations for the couplings in the spin sector are
\begin{subequations}
\begin{align}
	&\frac{dK_{s+}}{dl} = -\frac{K^2_{s+}}{8\pi^2} \left( g^2_{s1}+g^2_{s2}\right),\\
	&\frac{dK_{s-}}{dl} = \frac{1}{8\pi^2} g^2_{s2}-\frac{K^2_{s-}}{8\pi^2} g^2_{s1},\\
	&\frac{dg_{s1}}{dl} = (2-K_{s+}-K_{s-}) g_{s1}, \\
	&\frac{dg_{s2}}{dl} = (2-K_{s+} -\frac{1}{K_{s-}}) g_{s2}.
\end{align}
	\label{RG-equations}
\end{subequations}

\subsection{Luttinger Liquid Phase}
\label{sec:LL}

We start with the case where $g_{s1}=0$ and  $|g_{s2}(0)|$ small. The analysis for this regime is similar to what is done in the Kondo-Heisenberg chain (with nearest 
neighbor Kondo spins).\cite{zachar-2001,zachar-2001b,berg-2010} 

In this regime the RG equations of Eq. \eqref{RG-equations} become
 \begin{align}
 \begin{split}
	&\frac{dK^{-1}_{s+}}{dl} =\frac{dK_{s-}}{dl} = \frac{1}{8\pi^2} g^2_{s2},\\
	&\frac{dg_{s2}}{dl} = (2-K_{s+} -\frac{1}{K_{s-}}) g_{s2}.
 \end{split} \label{g1-reduced-eqns}
 \end{align}
To guarantee the $SU(2)$ invariance, these equations are subject to the initial conditions $K_{s-}(0) = K^{-1}_{s+}(0)\simeq1-g_{s2}(0)/4\pi$. 
The relation $K_{s-} = K^{-1}_{s+}$ is an invariant of this RG flow. Upon implementing this constraint, the set of RG equations \eqref{g1-reduced-eqns} can be further 
simplified to the  Kosterlitz RG equations
 \begin{align}
	\frac{dx}{dl}  = \frac{1}{8\pi^2}g_{s2}^2, \qquad  \frac{dg_{s2}}{dl} = 2x g_{s2},
 \end{align}
where $x=K_{s-}-1 \ll 1$. 
The solution of this flow equation is well known. Here we will only be interested in the $SU(2)$ invariant trajectories of the RG flow which satisfy $g=\pm 4\pi x$. Thus, 
a point on the $SU(2)$ invariant trajectory $x(0)=K_{s-}(0)-1 \simeq - g_{s2}(0)/4\pi < 0$ flows into the Gaussian  (free field) fixed point 
$g_{s1}=g_{s2}=0$, $K_{s+}=K_{s-}=1$ where the system is a (spin $1/2$) Luttinger Liquid which is a gapless and hence scale invariant system. 
In other words, the fixed point of $(g_{s1}=0,g_{s2}=\infty)$ is perturbatively unstable and flows to the Luttinger liquid fixed point along the  $SU(2)$ invariant RG 
trajectory.  
At this Luttinger liquid fixed point
all three different components of the triplet SC order parameter $\vec\Delta_a$ on the anti-bonding $a$-band and the Spin Density Wave (SDW) 
order on both bands decay as a power law. 
This is the Luttinger Liquid phase, shown in Fig.\ref{fig:SU(2)-RG}, and its correlators are standard. Since it has one gapless charge mode and two gapless spin modes it 
is a C1S2 Luttinger state (in the terminology of Balents and Fisher\cite{balents-1996}).

\subsection{The PDW phase}
\label{sec:PDW-phase}

In contrast, points along the other $SU(2)$ invariant trajectory, $x(0)=K_{s-}(0)-1 \simeq- g_{2}(0)/4\pi > 0$, flow to the strong coupling fixed point 
$(g_{s1}=0, g_{s2}=-\infty)$. 
At this fixed point both $\cos(\sqrt{4\pi}\theta_{s-})$ and $\cos(\sqrt{4\pi}\phi_{s+})$ acquire non-vanishing expectation values, the fields are pinned at values 
$(\theta_{s-},\phi_{s+})=\frac{\sqrt{\pi}}{2}( N_{s-},N_{s+}) $
(where $N_{s,\pm}$ are both odd or even integers at the same time)  and their quantum fluctuations are gapped.
It is easy to see that the identity $\ev{\cos{\sqrt{4\pi}\theta_{s-}}} = \ev{\cos{\sqrt{4\pi}\phi_{s+}}}$ holds along the $SU(2)$ invariant trajectories. This observation will be 
useful later. 

This means that a extended-Hubbard Heisenberg model on a  ladder when one of its bands is half-filled, has  is a Kosterlitz-Thouless (KT) phase transition from the 
gapless Luttinger Liquid  phase to the spin-gap phase described by the same strong coupling fixed point of the spin-gap phase (``Kondo singlet'') of the 
Kondo-Heisenberg chain (with nearest-neighbor Kondo ``impurities''). 

However, at this strong coupling fixed point observables that involve the dual fields of either $\phi_{s-}$ and $\theta_{s+}$ have vanishing expectation values and their 
correlations  decay 
exponentially fast. This results in short range correlations for the singlet uniform superconducting order parameter on the anti-bonding $a$ band, $\Delta_a $, and of the 
Charge Density Wave (CDW) order parameter, $O_\text{CDW}$ for both bands $a$ and $b$, which have the charge-spin factorized form
\begin{align}
	\Delta_a &=\frac{1}{\pi a} e^{-\im\sqrt{2\pi}\theta_{c,a}}\cos(\sqrt{2\pi}\phi_{s,a}),\nonumber\\
	O_{\textrm{CDW},a/b} &=\frac{1}{\pi a} e^{-\im\sqrt{2\pi}\phi_{c,a/b}}\cos(\sqrt{2\pi}\phi_{s,a/b})
\end{align}
where here $a$ is the lattice spacing.
For instance, in the case of the singlet uniform SC order parameter $\Delta_a$ for the anti-bonding $a$ band, in the $s\pm$ basis, its spin part is decomposed as 
 \begin{align}
 \cos(\sqrt{2\pi}\phi_{s,a})= &\cos(\sqrt{\pi}\phi_{s-})\cos(\sqrt{\pi}\phi_{s+})\nonumber\\
						+&\sin(\sqrt{\pi}\phi_{s-})\sin(\sqrt{\pi}\phi_{s+})
\end{align}
The presence of $\cos(\sqrt{\pi}\phi_{s-})$ and $\sin(\sqrt{\pi}\phi_{s-})$ guarantees exponentially decaying correlation functions. A similar form is also found  for the CDW 
operators of the bonding and anti-bonding bands. 

However this does not imply that there are no long range correlations at this fixed point since there are composite order parameters built from products of these operators 
which do have quasi long range order~\cite{zachar-2001}. To this end we define the order parameters for the PDW phase as the staggered part of the following product of 
operators in the two bands\cite{berg-2010}
\be
	O = \vec\Delta_a\cdot\vec S_b =\vec\Delta_a\cdot\vec J_b+(-1)^{x/a}O_\text{PDW} 
\ee
where $\vec S_b = \vec J_b + (-1)^{x/a}\vec N_b$, and $J_b$ is the total spin density vector on the bonding $b$ band and $\vec N_b$ is the N\'eel order parameter also 
for the bonding $b$ band. The explicit bosonized expression for the pair-density-wave order parameter $O_\text{PDW}$ is
\begin{align}
	&O_\text{PDW} = \vec\Delta_a\cdot\vec N_b=\frac{1}{2(\pi a)^2} \cos(\sqrt{2\pi}\phi_{c,b})e^{-\im\sqrt{2\pi}\theta_{c,a}}\nonumber\\
					&\qquad\times \left[2 \cos(\sqrt{4\pi}\theta_{s-}) +\cos(\sqrt{4\pi}\phi_{s-}) - \cos(\sqrt{4\pi}\phi_{s+})\right].
					\label{O-PDW}
\end{align}
It is easy to see that the spin part of $O_\text{PDW}$ has a nonzero expectation value in this phase. Therefore in spit of the fact that the correlation functions of the 
individual N\'eel (or SDW) order parameter of the bonding $b$ band and uniform triplet SC of the anti-bonding $a$ band are exponentially decaying, the correlation 
function of their product, the PDW order parameter  $O_\text{PDW}$, exhibits power law correlations:
 \be
 	\ev{O^{}_\text{PDW}(x)O^{\dagger}_\text{PDW}(0)} \sim {\cal C}^2_c {\cal C}^2_s |x|^{-2/K_{c,a}}
 \ee
in which ${\cal C}_c = \ev{\cos(\sqrt{2\pi}\phi_{c,b})}$ and ${\cal C}_s = \ev{\cos(\sqrt{4\pi}\theta_{s-})}$. Therefore $\Delta_\text{PDW}$ has quasi long range order with the 
exponent of $2K^{-1}_{c,a}$. The very same argument holds for a composite operator obtained from the product of the CDW order parameter on the bonding $b$ band 
and the uniform singlet SC on the anti-bonding $a$ band. Similar to the $O_\text{PDW}$ case, in $g_{s1}=0$ limit the individual CDW on the bonding $b$ band and 
uniform singlet SC on the anti-bonding $a$ band both decay exponentially fast but their product has quasi long range order with the same exponent as PDW order 
parameter. This is the Region II phase shown in Fig.\ref{fig:SU(2)-RG}.

\subsection{Uniform Superconducting Phase}
\label{sec:uniform-sc-phase}

Let us now consider the  opposite limit of $g_{s2}=0$.  Here one can   repeat a similar RG analysis as in the $g_{s1}=0$ limit (or make use of  the duality symmetry hidden 
in the problem, shown in Eq.\eqref{eq:duality}) to show that the RG with $0 < g_{s1}(0) \ll 1$ will be renormalized back to $g_{s1}=0$ along the $SU(2)$ invariant 
trajectory and we are again in the Luttinger Liquid phase we found before. Therefore, just like the fixed point at $(g_{s1}=0,g_{s2}=\infty)$ in the former regime, the  fixed 
point $(g_{s1}=\infty, g_{s2}=0)$ is also not accessible in this case. This means that components of the uniform triplet SC or the N\'eel (SDW) order parameters have 
power law correlations. 

However, assuming again $SU(2)$ invariance, in a system with $g_{s2}=0$ a small negative $g_{s1}<0$ will flow to the strong coupling fixed point 
$(g_{s1}=-\infty,g_{s2}=0)$. This is the fixed point dual to the $(g_{s1}=0,g_{s2}=-\infty)$ fixed point under the duality transformation of Eq.\eqref{eq:duality}. 
In this phase now the expectation values $\ev{\cos(\sqrt{4\pi}\phi_{s\pm})}$ are nonzero and observables that are 
functions of $\phi_{s\pm}$ have long-ranged correlations. Instead the expectation values  $\ev{\cos(\sqrt{4\pi}\theta_{s-})}=0$ and its fluctuation 
has short-ranged correlations. At this strong coupling fixed point the semi-classical expectation values of the $\phi_{s,a}$ and $\phi_{s,b}$ are such that 
$\ev{\cos(\sqrt{4\pi}\phi_{s+})}\times \ev{\cos(\sqrt{4\pi}\phi_{s-})} >0$ and hence $\sqrt{4\pi}\phi_{s+}=\sqrt{4\pi} \phi_{s-}=0,\pi \mod{2\pi}$. 
In a phase controlled by this fixed point the two sectors, $\pm$, have separate spin gaps. 
Furthermore, in this regime  the expectation value $\langle \cos(\sqrt{2\pi}\phi_{s,a})\rangle \neq 0$ is nonzero, and therefore in the phase controlled 
by this fixed point the uniform singlet SC on the anti-bonding $a$ band has quasi long range order,
\be
	\ev{\Delta^{}_a(x)\Delta_a^\dagger(0)} \sim {\cal C}^2_{s,a} |x|^{-2/K_{c,a}}.
\ee
where ${\cal C}_{sa} = \ev{\cos(\sqrt{2\pi}\phi_{s,a})}$ in this phase. By the same argument, the CDW order parameter of the anti-bonding $a$ band has 
quasi long range order as well,
\be
	\ev{ O^{a}_\text{CDW}(x){O^{a}}_\text{CDW}^\dagger(0)} \sim {\cal C}^2_{s,a} |x|^{-2K_{c,a}},
\ee
However, given the repulsive nature of the interactions the Luttinger parameters obey  $K_{c,a}<1$ and therefore SC is the dominant fluctuations of this phase. 
Nevertheless, unlike the PDW phase, this phase has dominant uniform SC correlations. Nevertheless we note that in this phase there exists subdominant 
CDW correlations.

On the other hand we will now see that the correlation function of $O_\text{PDW}$ decays exponentially fast in this phase. Similar to the previous discussion, it is easy to 
see that $\ev{\cos(\sqrt{4\pi}\phi_{s+})} = \ev{\cos(\sqrt{4\pi}\phi_{s-})}$  for a SU(2) invariant RG flow. 
Therefore, looking back at the structure of the PDW order parameter given in Eq.\eqref{O-PDW}, we see  that the expectation values of $\cos(\sqrt{4\pi}\phi_\pm)$ cancel 
each other out at this fixed point. Hence, the expectation value of the spin part of $O_\text{PDW}$ is zero since  the expectation value $\ev{\cos(\sqrt{4\pi}\theta_{s-})} =0$ 
vanishes exactly in this phase. Moreover the two-point correlation function of $O_\text{PDW}$ has to be proportional to  the vertex operator $ \cos(\sqrt{4\pi}\theta_{s-})$ 
whose correlations decay exponentially fast. Consequently the correlations of $O_\text{PDW}$ are short-ranged. 

On the other, since there are independent spin gaps on both bands, the product of SC on $a$-band and CDW on $b$-band still has long range order . Therefore this product has similar correlations at both phases. This means it can not be used as an order parameter to distinguish these two phases. Therefore $O_\text{PDW}$ is the unique order parameter to distinguish the state with PDW order from the state with uniform SC order at the spin-gapped regime of the two-leg ladder system with one band kept at half-filling. This is the Region I phase shown in Fig.\ref{fig:SU(2)-RG} and it is identified with a phase with uniform superconductivity.

\subsection{The PDW-uniform SC Quantum Phase Transition}
\label{sec:PDW-phase-transition}

We will now discuss the quantum phase transition between the state with uniform SC order (with power law correlations) and the PDW state. To this end we first note that 
both PDW and uniform SC in their associate phases have quasi long range order with the same exponent of $2/K_{c,a}$. This happens since the exponents are 
controlled in both cases by the decoupled charge degree of freedom left in the system. However  the two phases are distinguished by the  fact that one SC state is 
staggered (the PDW state) while the other is uniform. Therefore by symmetry the phase transition between these two states is similar to the transition from a state with 
translational symmetry to the one with only a $\mathbb{Z}_2$ discrete broken translational symmetry. So we expect to find that this transition to be in the Ising universality 
class. However, as we will see, the way this happens is actually rather subtle.

To discuss the nature of the phase transition between Region I and Region II of Fig.\ref{fig:SU(2)-RG}  we need to look at the $g_{s1}=g_{s2}$ line in the parameter 
space. The duality symmetry of the Hamiltonian implies that the phase diagram must be symmetric under it. We will now see that there is a direct quantum phase 
transition at the self-dual line, the half-line that separates Region I from Region II in Fig.\ref{fig:SU(2)-RG}. 
 From the first  RG equation in Eq.\eqref{RG-equations}, it is clear that  the Luttinger parameter $K_{s+} \rightarrow0$ flows to zero whenever either $g_{s1}$ or $g_{s2}$ 
 is relevant. Furthermore along the $g_{s1}=g_{s2}$ line system flows to a new strong coupling fixed point with  $\ev{\cos(\sqrt{4\pi}\phi_{s+})}\neq0$ and the field 
 $\phi_{s+}$ is pinned. At the strong coupling fixed point the effective bosonized Hamiltonian can be further simplified to the following
\begin{align}
 	{\cal H}_\text{eff} =& \frac{v_{s-}}{2} \left\{ K^{}_{s-}(\partial_x\theta_{s-})^2+K_{s-}^{-1}(\partial_x\phi_{s-})^2 \right\}\nonumber\\
		&+\frac{\mu_\phi}{\pi} \cos(\sqrt{4\pi}\phi_{s-}) + \frac{\mu_\theta}{\pi} \cos(\sqrt{4\pi}\theta_{s-}), \label{eq:effective-hamiltonian}
\end{align}
where the new couplings $\mu_{\phi/\theta}$ are related to the $g_{s}$'s as
\be
	\mu_{(\phi,\theta)} = \frac{g_{s(1,2)}}{2\pi }  \; \ev{\cos(\sqrt{4\pi}\phi_{s+})}.
\ee
This system was discussed extensively by Lecheminant et.~al.\cite{lecheminant-2002}, who showed at $K_{s-}=1$ the resulting $H_\text{eff}$ of 
Eq.\eqref{eq:effective-hamiltonian} can be re-fermionized in terms of two sets of chiral Majorana fermions. For $\mu_\theta=\mu_\phi$, {\it i.e.} along the self-dual line of 
our problem,  one of the chiral Majorana fermion pairs becomes massless and its mass changes sign accross this phase transition. Therefore, the phase boundary 
between the phases corresponding to $g_{s1}$ and $g_{s2}$ is the in the same universality class as the quantum Ising chain, a theory of a (non-chiral) massless 
Majorana fermion. We should point out that the expression of the Ising order (and disorder) operators in terms of the bosonized fields of the ladder is highly non-local.

This question can also be addressed directly through RG equations as well. Indeed along the self-dual line they reduce to
\begin{align}
	\begin{split}
	&\frac{dK_{s-}}{dl} = \mu_\theta^2-K_{s-}^2\mu_\phi^2 ,\\
	&\frac{d\mu_\phi}{dl} = (2-K_{s-}) \mu_\phi, \\
	&\frac{d\mu_\theta}{dl} = (2-\frac{1}{K_{s-}}) \mu_\theta.
	\end{split}
 \end{align} 
Starting with $K_{s-}(0)=1$ and $|\mu_\theta(0)|>|\mu_\phi(0)|$, we see that RG flows to the the fixed point of $(|\mu_\theta| =\infty, \mu_\phi=0, K_{s-}=\infty)$ which is the 
same as the PDW fixed point. In contrast, if we start with $K_{s-}(0)=1$ and $|\mu_\phi(0)|>|\mu_\theta(0)|$, the RG flow will take us to 
$(\mu_\theta=0,| \mu_\phi|=\infty, K_{s-}=0)$, i.e. the uniform SC fixed point. The flow with the initial condition $K_{s-}(0)=1$ and $g_{s1}(0)=g_{s2}(0)<0$ will
 flow to the Ising critical point with $g_{s1}=g_{s2}=-\infty$ and $K_{s-}=1$ while $g_{s1}=g_{s2}>0$ flows to the Gaussian fixed point.

Fig.\ref{fig:SU(2)-RG} shows the results of a numerical calculation of all the $SU(2)$-invariant RG flows projected onto the $(g_{s1},g_{s2})$ plane. Solid black lines 
represent the separatrix between different phases. The low energy behavior of all the models with $SU(2)$-invariance which satisfy $g_{s1},g_{s2}>0$ is controlled by the 
Gaussian fixed point. The rest of the flow, except for the semi line $g_{s1}=g_{s2}<0$, will end either at $(g_{s1}, g_{s2})=(-\infty,0)$ or at $(g_{s1},g_{s2})=(0,-\infty)$ 
depending on the initial values of the couplings. The RG flow pattern is symmetric around the $g_{s1}=g_{s2}$ line as dictated by the duality symmetry.


\section{Bonding band at general commensurate fillings}
\label{sec:other-commensurabilities}

We will now discuss the case of a ladder with a bonding band at other commensurate fillings.
To understand this case it is useful to recall first the physics of the simpler problem of the extended repulsive one-band Hubbard-Model at quarter filling $k_F = \pi/4$. 
This system has a quantum phase transition at a critical value of the nearest  neighbor ``Coulomb'' interaction between a Luttinger liquid and a commensurate  
insulating CDW state. An intuitive classical picture of the ground state of such a system  is as if the electrons occupy every other site on the lattice 
with their spins arranged in a ``stretched'' N\'eel  antiferromagnetic state, as in an antiferromagnetic Heisenberg chain with twice the lattice constant. 
In this regime, the charge sector of the Hubbard chain is gapped (and hence insulating).

This phase transition is driven by a higher order Umklapp interaction, which appears at third order in perturbation theory, which stabilizes this period $2$ CDW state. 
As it is well known, and easy to see, in the bosonized for of this system the Umklapp term for the $1/4$ filled band has the form  
(see, {\it e.g.} Ref.[\onlinecite{giamarchi-2003}] and references therein)
\begin{equation}
{\cal H}_{u,1/4}=g_{1/4} \cos(4\sqrt{2\pi}\phi_c )
\end{equation}
The scaling dimension of this Umklapp operator is $4K_c$, where $K_c$ is the charge Luttinger parameter of the extended Hubbard chain. 
Therefore, this Umklapp process  is relevant for $K_c < 1/4$ which always lays in the intermediate to strong coupling repulsive regime. 
Although in this regime the bosonization formulas that relate the {\em parameters} of the microscopic model (in its naive continuum limit) 
and the bosonized theory is no longer accurate, the {\em form} of the effective low energy bosonized theory retains its form as it is dictated by symmetry. 
The main problem is that the connections between the Luttinger parameter(s) and the microscopic parameters is more complex due to finite renormalizations 
induced by the irrelevant operators. In practice this relation must (and is) determined from numerical calculations on the microscopic model. 

For a system on a two-leg ladder one can pursue a similar line of reasoning and use the fact that, for certain fillings,  there is a similar Umklapp processes for  the bonding 
band (for instance). However, just as in the case of the extended Hubbard chain, here too 
the couplings corresponding to these Umklapp terms in the strong coupling effective theory can not be easily related to the microscopic parameters of 
the original lattice model and requires serious numerical work. Nevertheless, such Umklapp processes  still exist and should eventually become relevant. 
At this value of the parameters, where $K_c=1/4$, the Umklapp process for the bonding band becomes marginally relevant, and the system has a 
Kosterlitz-Thouless phase transition to a period $2$ commensurate CDW state that coexists with antiferromagnetic order, with power law correlations in the spin sector. 

We can now follow the same approach we used for the case of a half-filled bonding band of the preceding sections to determine the phase diagram. We will not present 
the details here but give the main results. As before, the phase diagram in general has three phases: a) a Luttinger Liquid phase (similar to the one discussed by 
Wu {\it et al}\cite{wu-2003}), b) a phase with uniform superconducting order (and hence a spin gap), and c) a PDW phase (also with a spin gap).  
However, the ordering wave vector of the PDW state is now $Q_\text{PDW}=\pi/2$ (instead of $Q_\text{PDW}=\pi$ for the case discussed in the preceding sections). 
This  PDW state there is a  ``composite'' CDW quasi ordered state (with power-law correlations)  with degrees of freedom partially on the anti-bonding band, 
in this case with wave vector $Q_\text{CDW}^a=2k_{F,a}+Q_\text{PDW}$. 
Thus the resulting PDW phase  is similar to the one discussed in  Refs. [\onlinecite{zachar-2001b,berg-2010}].  
In contrast, the phase with uniform SC order has a CDW state that develops on the anti-bonding bans alone and has a conventional $2k_{F,a}$ 
ordering wave vector. In spite of these differences with the case of the half-filled bonding band, the quantum phase transition between the PDW phase 
and the phase with uniform SC order for the quarter filled bonding band is also in the Ising universality class.

A state with very similar properties was found in the Kondo-Heisenberg chain for a Kondo lattice with period $2$ (see Ref.[\onlinecite{berg-2010}]). However, while in the 
KH chain translation invariance is broken explicitly by the assumed fixed spacing of the Kondo spins, in the case of the ladder translation invariance is broken 
spontaneously at the Kosterlitz-Thouless quantum phase transition we just discussed. However the spin gap (and the PDW state as well) can only develop once this 
CDW state is formed. In this sense this is an example of {\em intertwined} (as opposed to {\em competing}) orders in the sense discussed by 
Berg {\it et al}.\cite{berg-2007,berg-2009b}

This line of argument can, naturally, also be extended to states in which the bonding band has other fillings, such as $1/2^{n}$ for $n=1,2,...$, and consider Umklapp 
terms of the form $g_{\frac{1}{2n}}\cos(2n\sqrt{2\pi}\phi_c)$. Although such terms will generally be present, their effects become relevant only for $K_c<1/n^2$ which lays 
deep in in the extreme repulsive regime for $n\geq2$. Thus, unless the system has substantial interactions beyond nearest neighbors, the resulting higher order 
commensurate CDW and PDW states will be quite weak and difficult to stabilize.

\section{PDW state in an extended Hubbard-Heisenberg model on a  two-leg ladder with $\Phi$ flux per plaquette}
\label{sec:flux}

\begin{figure}[hbt]
	\begin{center}
	\subfigure[]{\includegraphics[width=0.35\textwidth]{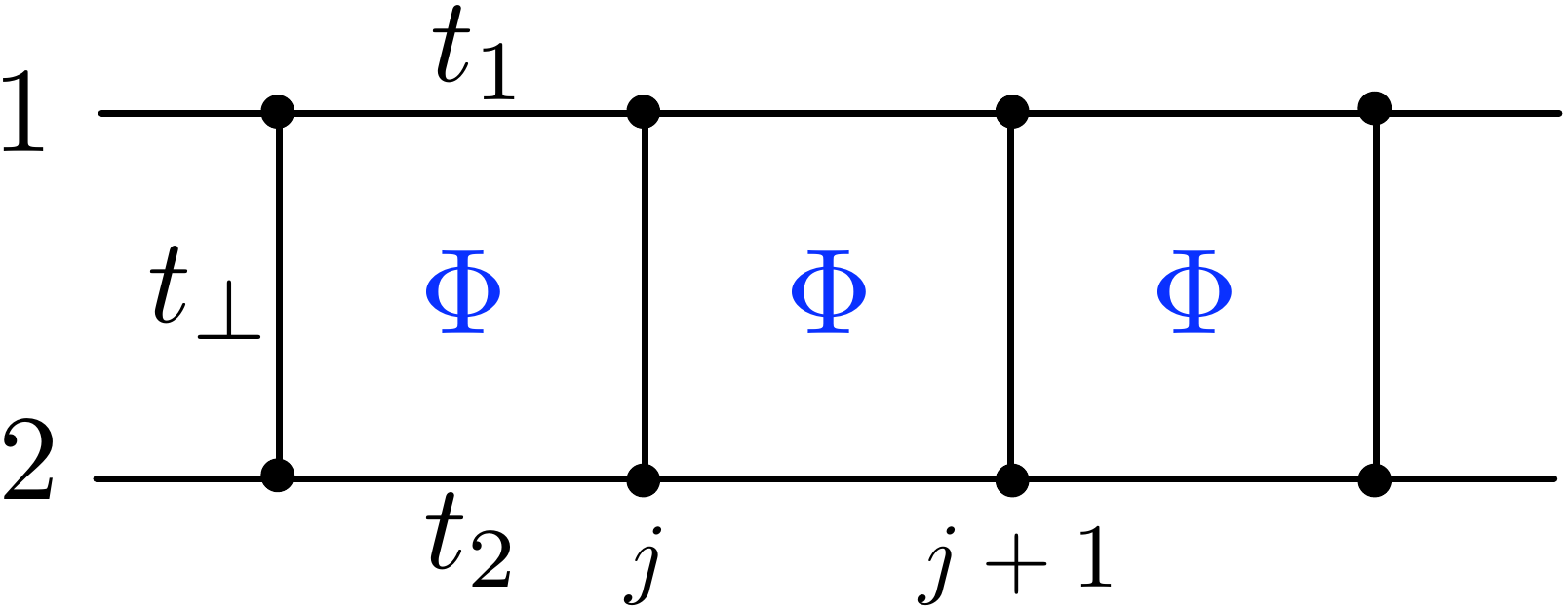}}
	\subfigure[]{\includegraphics[width=0.4\textwidth]{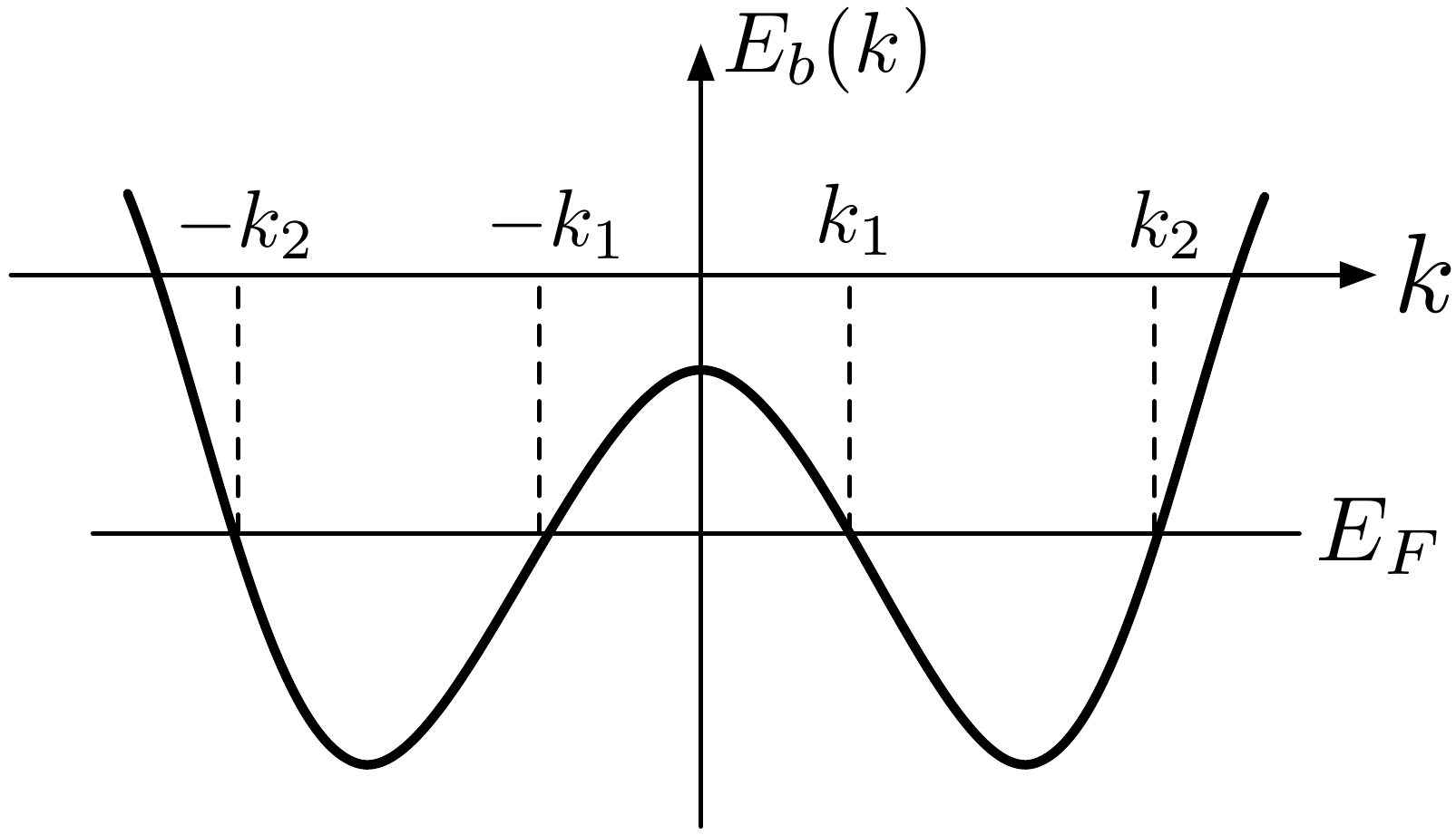}}
	\end{center}
	\caption{a) A two-leg ladder with flux $\Phi$ (in units of the flux quantum) per plaquette. Here $1$ and $2$ label the two legs, $j$ and $j+1$ label two consecutive rungs. The hopping amplitudes are $t_1= t\;  e^{\im\Phi/2}$, $t_2=t \; e^{-\im \Phi/2}$ and $t_\perp$. 
	b) Schematic plot of the dispersion relation of the bonding band $E_b(k)$ for a non-vanishing flux per plaquette, $\Phi\neq 0$.}
	\label{fig:double-minima}
	\end{figure}

We will now introduce and investigate another ladder model in which we can show has a PDW phase.  More specifically we will consider an extended Hubbard-
Heisenberg model in a two-leg ladder with flux $\Phi$ per plaquette (in units of the flux quantum $\phi_0=hc/e$).
As usual the flux is introduced by a Peierls substitution which here reduces to assigning a phase $\pm \Phi/2$  to the electron hopping matrix elements along the two legs 
which now become complex and are complex conjugate of each other,
$t^{}_1=t^*_2$, where $t_{1,2}$ are the hopping amplitudes on top and bottom leg (see Fig.\ref{fig:double-minima} a).
In addition to the the hopping along the rungs, we assume a real hoping amplitude between the legs.

The free part of the Hamiltonian of this system is
\begin{align}
	H_0=&-t \sum_j \left( e^{\im\Phi/2}c^\dagger_{1,j+1} c^{}_{1,j} + e^{-\im\Phi/2}c^\dagger_{2,j+1} c^{}_{2,j} + \text{h.c.}\right)\nonumber \\
		& -t_\perp\sum_j \left( c^\dagger_{1,j} c^{}_{2,j} +c_{2,j}^\dagger c_{1,j}\right)
\end{align}
in which $i=1,2$ is the chain index referring to the top and bottom chains respectively and $j\in {\mathbb Z}$ denotes the lattice sites. To the best of our knowledge this 
electronically frustrated system has not been discussed previously. The interaction terms, $H_{\rm int}$ that we will consider are the same as in the conventional ladder 
system and are given in Eq.\eqref{lattice-model}. 

We will see here that this model has a very rich phase diagram which we will not explore completely. However we will show that PDW phases occur naturally although 
through a rather different mechanism that we found in the conventional ladder discussed in the previous sections.
We will discuss this problem using the same methods bosonization we used above. 

We begin by constructing an effective field theory. In momentum space the free fermion part of the Hamiltonian becomes
\begin{widetext}
\begin{equation}
	H_0 = \int_{-\pi}^\pi \frac{dk}{2\pi} \Big[-2t \cos(k+\Phi/2)c^\dagger_1(k)c^{}_1(k)
	 -2t \cos(k-\Phi/2)c^\dagger_2(k)c^{}_2(k)
	-t_\perp( c^\dagger_1(k)c^{}_2(k) +c_2^\dagger(k)c_1(k))\Big]
\end{equation}
\end{widetext}

For  $t_\perp=0$ the band structure consists of right and left  branches centered at $k=\Phi/2$ and $k=-\Phi/2$ respectively. For $t_\perp\neq 0$ a full gap opens up at the 
crossing point of the two bands and bonding and anti-bonding bands form.

The Hamiltonian is diagonalized if we switch to the new basis defined by the orthogonal transformations $c_\eta (k)= \sum_{i} M_{\eta,i}(k) c_i(k)$ with the 
transformation matrix $M(k)$: 
\be
	M_{\eta,i}(k)=
		\begin{pmatrix}
			\cos(\xi(k)/2)& -\sin(\xi(k)/2)\\
			\sin(\xi(k)/2)& \cos(\xi(k)/2)
		\end{pmatrix}
\ee
where $\eta=a,b$ stands for the bonding and anti-bonding bands, $i=1,2$ labels the legs of the ladder, and $\xi$ is defined as
\begin{align}
	\sin(\xi(k)/2) =& \frac{u(k)}{\sqrt{1+u^2(k)}}\nonumber\\
	\cos(\xi(k)/2) = &\frac{1}{\sqrt{1+u^2(k)}}
\end{align}
in which
\be
	t_\perp u(k) =2t\sin(\Phi/2)\sin(k) + \sqrt{(2t\sin(\Phi/2)\sin(k))^2+t^2_\perp}.
\ee
The inverse transformations are defined by the inverse matrix $M^{-1}_{i,\eta}$ as following
\begin{align}
	c_{1,j,\sigma} &=\int^\pi_{-\pi} \frac{dk}{2\pi} e^{\im j k}\left[ \cos(\frac{\xi(k)}{2})c^a_{\sigma}(k) + \sin(\frac{\xi(k)}{2})c^b_{\sigma}(k)\right]\nonumber\\
	c_{2,j,\sigma}&=\int^\pi_{-\pi} \frac{dk}{2\pi} e^{\im j k}\left[- \sin(\frac{\xi(k)}{2})c^a_{\sigma}(k) +\cos(\frac{\xi(k)}{2})c^b_{\sigma}(k)\right]
	\label{eq:bonding-antibonding-lattice-transformation}
\end{align}
where $b$ and $a$ label the bonding and the anti-bonding bands respectively.

The dispersion relations for the bonding and anti-bonding bands are
\be
	E_\eta(k) = -2t\cos(\Phi/2)\cos(k) \pm \sqrt{(2t\sin(\Phi/2)\sin(k))^2+ t^2_\perp}.
	\label{eq:band-structure-flux}
\ee
Band structures of this type appear in quantum wires with two pockets~\cite{datta-2009} and in 1D electronic systems with spin-orbit interactions (with the leg label playing the role of the electron spin) (see, {\it e.g.} Ref.[\onlinecite{lutchyn-2011}] and references therein).
The dispersion relations  $E_\eta(k)$ satisfies the symmetries
\begin{equation}
E_\eta(-k)=E_\eta(k), \qquad E_a(\Phi+2\pi,k) = - E_b(\Phi,k)
\end{equation}
For a wide range of parameters, $t_\perp/t$ and flux $\Phi$, the bonding band  has the  form sketched in Fig.\ref{fig:double-minima} with two minima in momentum space. For the rest of this section we will focus on the regime of the parameters in which the Fermi energy lies below the hybridization gap of the bonding band, $E_F<E_b(0)=-2t \cos(\Phi/2)-t_\perp$. In this regime, the Fermi energy crosses the bonding band at four distinct points, $\pm k_1, \pm k_2$, while the anti-bonding band is empty, as shown in Fig. \ref{fig:double-minima} b.

Here we will consider an extended Hubbard-Heisenberg model on a ladder with flux $\Phi$ per plaquette. We will see now that this system has an interesting phase structure. We will analyze this system using the same bosonization methods as in the conventional ladder. For reasons that we will explain below we will focus on the spacial case of flux $\Phi=\pi$ per plaquette.

\subsection{Low energy continuum limit}
\label{sec:low-energy-pi-flux}

In this work we are interested in the regime in which the Fermi energy crosses only the bonding band and hence the anti-bonding band is empty.  Furthermore we assume 
that the interactions are weak enough that we can focus only in the low energy excitations near the four Fermi points $\pm k_1, \pm k_2$ where the Fermi energy crosses 
the bonding band. Furthermore, for the more interesting case of flux $\Phi=\pi$, the Fermi points obey the commensurability condition $k_1+k_2=\pi$. In this case we also 
have $u(k_1)=u(k_2)$. Hence the parameter $\xi(k)$ obeys the same identity and it is the same for both Fermi points. Henceforth it will be 
 denoted by $\xi$.
 
By looking only at the low energy fluctuations around the Fermi points in the bonding band, the expansion of Eq.\eqref{eq:bonding-antibonding-lattice-transformation} 
reduces to the operator identifications 
\begin{subequations}
\begin{align}
	 \frac{1}{\sqrt{a}} c_{1,\sigma}(j) &\to
	  \sin(\frac{\xi}{2})L_{1\sigma}(x)e^{\im k_1x}  + \cos(\frac{\xi}{2})R_{1\sigma}(x)e^{-\im k_1x} \nonumber\\
	&+ \sin(\frac{\xi}{2}) R_{2\sigma}(x)e^{\im k_2x} + \cos(\frac{\xi}{2})L_{2\sigma}e^{-\im k_2x}\\
	 \frac{1}{\sqrt{a} } c_{2,\sigma}(j)& \to 
	 \cos(\frac{\xi}{2})L_{1\sigma}(x) e^{\im k_1x} + \sin(\frac{\xi}{2})R_{1\sigma}(x)e^{-\im k_1x}  \nonumber\\
	&+ \cos(\frac{\xi}{2}) R_{2\sigma}(x)e^{\im k_2x}+ \sin(\frac{\xi}{2}) L_{2\sigma}(x)e^{-\im k_2x}
\end{align}
	\label{eq:lattice-to-bonding-transformations}
\end{subequations}
where we have used $\cos(\xi(-k)/2) = \sin(\xi(k)/2)$ and $\xi(k_1)=\xi(k_2)\equiv \xi$ which are both true for $\Phi=\pi$, and where we have also projected out the anti-
bonding band.
We will treat the Fermi point labels as a flavor index $f=1,2$.

By inspection of the free fermion lattice Hamiltonian one can see that the Fermi momenta $k_1$ and $k_2$ are essentially determined by the flux $\Phi$ and by the filling 
fraction of the bonding band. In what follows we will ignore the contribution of the anti-bonding band since it is empty and its excitations have energies larger than the 
cutoff of the effective field theory.

Following a similar discussion as in section \ref{sec:model}, the non-interacting continuum Hamiltonian becomes
\be
	{\cal H}_0= \sum_{f=1,2} (-\im v_f)\left\{R^\dagger_{f,\sigma}\partial_x R^{}_{f,\sigma} - L^\dagger_{f,\sigma} \partial_x L^{}_{f,\sigma}\right\}
\ee
where $v_1=-\frac{dE_b}{dk}|_{k_1}$ and $v_2=\frac{dE_b}{dk}|_{k_2}$ are the Fermi velocities associated with the two Fermi points. For general flux $\Phi$ there is no 
symmetry relation the Fermi points and the two Fermi velocities are different, $v_1\neq v_2$. 

However, the the case of flux $\Phi=\pi$ per plaquette the energy bands have the additional symmetry $k \to \pi -k$. This symmetry reflects that fact that an exchange of 
the two legs, $1 \leftrightarrow 2$, is in general equivalent to the reversal of the flux $\Phi \leftrightarrow -\Phi$ which is the time-reversed state. However due to the flux 
quantization, the states with $\Phi=\pi$ and $\Phi=-\pi$ are equivalent since the Hamiltonian is a periodic function of the flux with period $2\pi$ (corresponding to 
unobservable changes  by an integer multiple of the flux quantum).  On the other hand, from  Eq.\eqref{eq:band-structure-flux}, we see that for flux $\Phi=\pi$ the 
dispersion relations are also invariant under $k \to \pi - k$ which amounts to an exchange of the two fermi points. 
Thus, in the case of flux $\Phi=\pi$   which insures that the Fermi velocities are equal, $v_1=v_2$, for all fillings of the bonding band (and of the anti-bonding band as 
well). Therefore, for flux $\Phi=\pi$, the symmetry of exchanging the two legs implies that the effective low energy theory must have a symmetry under the exchange of the 
flavor labels $1$ and $2$ (together with a chiral transformation which exchanges right and left movers).

\begin{figure}[t!]
		\includegraphics[width=0.35 \textwidth]{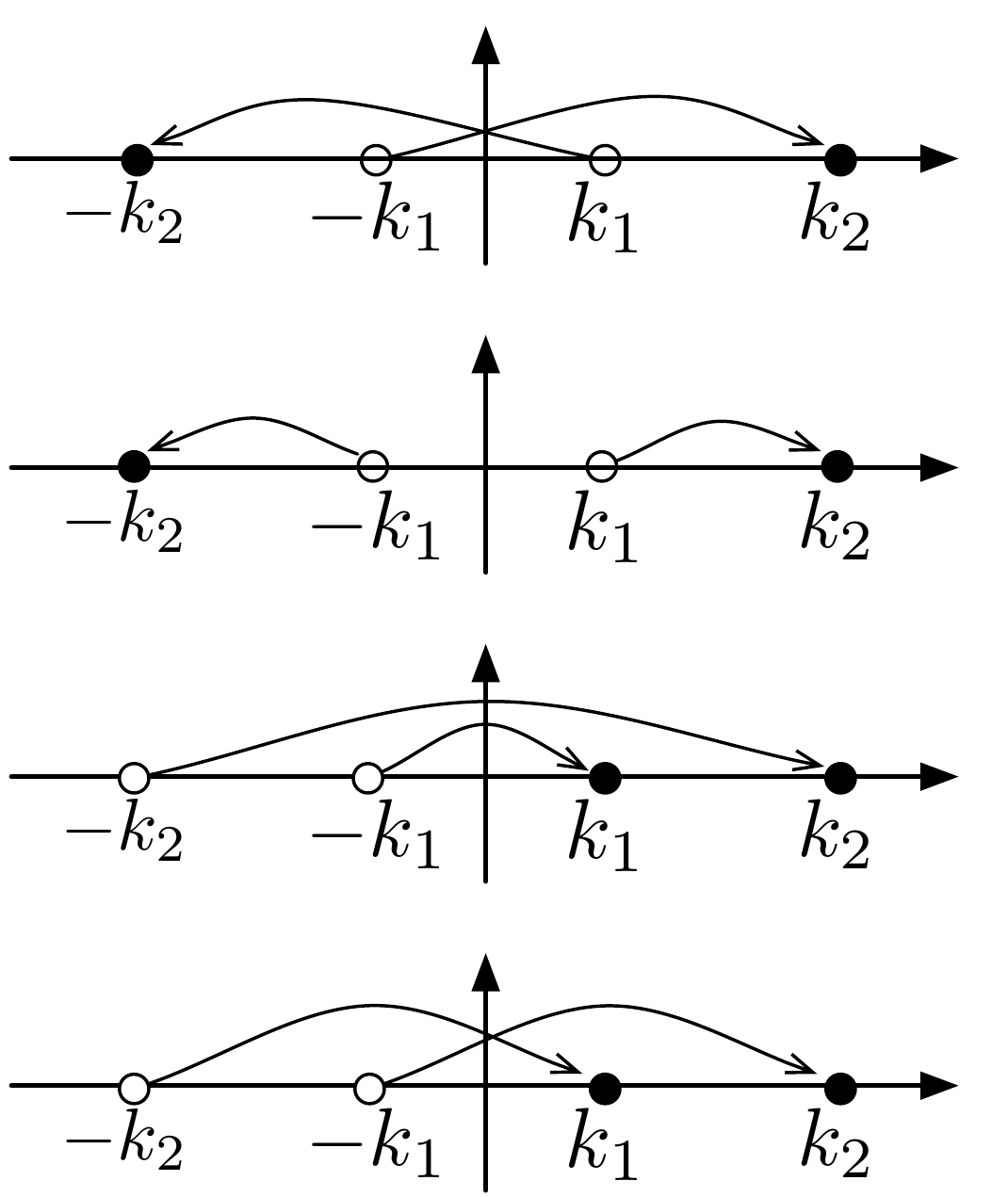}
		\caption{Schematic representation of all the processes leading to uniform SC couplings and $Q=2(k_1+k_2)$ Umklapp processes. The sum of the top two 
		diagrams represent the uniform singlet and triplet SC interactions while the two lower diagrams correspond to the Umklapp process.}
		\label{fig:interaction-diagram}
	\end{figure}

In order to introduce all possible four-fermion intra-band and inter-band interactions, one considers an extended Hubbard-Heisenberg type lattice problem, just as what 
we did for the two-leg ladder system of Section~\ref{sec:model}, and construct the continuum theory for the present case. All four-fermion interactions for this system  can 
be represented by simple diagrams similar of the type shown in Fig.\ref{fig:interaction-diagram}. All the interactions can again be classified into charge and spin current-
current interactions, singlet and triplet SC couplings or Umklapp processes with different commensurabilities. 

This means that the effective field theory for the present system has the same field theoretical form as the Hamiltonian of the two-leg ladder system given by 
Eq.~\eqref{smooth-continuum-H}. The only difference is that  in the present case, the two sets of right- and left-moving labeled by the flavor index $f=1,2$ are low energy 
fluctuations of the bonding band. Moreover the connection between the couplings in the effective theory and the microscopic parameters of the original lattice problem is 
different for the two problems. The two top diagrams in Fig.~\ref{fig:interaction-diagram} represent singlet and triplet SC interactions between the $1$ and $2$ while the 
lower diagrams corresponds to the most relevant $Q=2(k_1+k_2)$ Umklapp processes. We will further assume that the Fermi points $\pm k_1, \pm k_2$ are such that no 
other Umklapp processes are allowed.

The discussion of the phase diagram of this system in the incommensurate regime is analogous to what has been discussed in the conventional ladder by many authors. 
C. Wu {\emph et. al.}\cite{wu-2003} find that the only SC state in the phase diagram of the system away from the half-filling is the uniform $s$- or $d$-wave SC. Similar 
conclusions hold for the incommensurate regime of the this model which is in Luttinger liquid phase with two gapless charge modes and two gapless spin modes, 
$C2S2$. The only difference is that in this case these modes originate entirely from the bonding band.

For general flux $\Phi$ and for certain filling fractions of the bonding band Umklapp process involving separately the pairs of Fermi points $\pm k_1$ and $\pm k_2$ 
become allowed. The physics that follows in these cases is similar to what we discussed for the conventional ladder in Section \ref{sec:model}  and will not be repeated 
here. 

However, for flux $\Phi\neq 0$ a new type of Umklapp process, shown in Fig.\ref{fig:interaction-diagram}, becomes allowed.  For this process to be possible Fermi 
momenta must satisfy the condition $Q=2(k_1+k_2)$. This Umklapp process leads to the following interactions: 
\begin{align}
	{\cal H}_\text{Um} = \left( \lambda_{u3} \, n^\dagger_1n^{}_2 + \lambda_{u4} \, \vec n^\dagger_1\cdot \vec n^{}_2 \right) e^{\im Qx} + \text{h.c.}
\end{align}
where $n_f$ (with $f=1,2$) are the $2k_F$  CDW order parameters associated with the Fermi points at $\pm k_1$ and $\pm k_2$, with ordering wave vectors $2k_1$ and 
$2k_2$ respectively. Similarly,  $\vec n_f$ is the associated  SDW order parameters with the same ordering wave vectors. When the commensurability condition is 
satisfied this process is marginal and needs to be included in the effective low energy theory.

However the commensurability condition $k_1+k_2=\pi$ can only be met if the flux is $\Phi=\pi$. Furthermore, in this case  the system is commensurate for all fillings of 
the bonding band. 
For $\Phi=\pi$ the one-particle spectrum is given by $E_b(k)=-\sqrt{4t^2\sin^2(k)+t^2_\perp}$ which satisfies $E_b(\pi-k)=E_b(k)$. Therefore if $k_1$ is a Fermi 
momentum so is $k_2=\pi-k_1$. Hence for flux $\Phi=\pi$ the system remains commensurate {\em for all}  electron fillings. 
We will see below that for $\Phi=\pi$ the pair-density-wave state exists for all values of the filling (with the Fermi energy in the bonding band). 
From now on we will restrict ourselves to the case of flux $\Phi=\pi$. 

The bosonized Hamiltonian for flux $\Phi=\pi$ is (including the Umklapp process) 
\begin{align}
	{\cal H} =&  \frac{v_{c+}}{2} \left\{ K_{c+} (\partial_x\theta_{c+})^2 + K^{-1}_{c+} (\partial_x\phi_{c+})^2 \right\}\nonumber\\
		       +& \frac{v_{c-}}{2} \left\{ K_{c-} (\partial_x\theta_{c-})^2 + K^{-1}_{c-} (\partial_x\phi_{c-})^2 \right\}\nonumber\\
		       +&\frac{v_{s+}}{2} \left\{ K_{s+} (\partial_x\theta_{s+})^2 + K^{-1}_{s+} (\partial_x\phi_{s+})^2 \right\}\nonumber\\
		       +& \frac{v_{s-}}{2} \left\{ K_{s-} (\partial_x\theta_{s-})^2 + K^{-1}_{s-} (\partial_x\phi_{s-})^2 \right\}\nonumber\\
		       + &\frac{\cos(\sqrt{4\pi}\phi_{s+})}{2(\pi a)^2} \left[ g_{s1} \cos(\sqrt{4\pi}\phi_{s-}) + g_{s2} \cos(\sqrt{4\pi}\theta_{s-})  \right]\nonumber\\
		       +&  \frac{\cos(\sqrt{4\pi}\phi_{s+})}{2(\pi a)^2}   \left[g_{5}  \cos(\sqrt{4\pi}\theta_{c-})+g_{u5} \cos(\sqrt{4\pi}\phi_{c-}) \right]\nonumber\\
		      + &\frac{\cos(\sqrt{4\pi}\theta_{c-})}{2(\pi a)^2} \left[ g_{3} \cos( \sqrt{4\pi} \theta_{s-} ) + g_{4} \cos(\sqrt{4\pi} \phi_{s-})\right] \nonumber\\
		       +& \frac{\cos(\sqrt{4\pi}\phi_{c-})}{2(\pi a)^2} \left[ g_{u3} \cos( \sqrt{4\pi} \theta_{s-} ) + g_{u4} \cos(\sqrt{4\pi} \phi_{s-}) \right]
\label{eq:Heff}
\end{align}
where $\phi_\pm =(\phi_2\pm \phi_1)/\sqrt{2}$ and similarly for $\theta$ fields. 
As before, there are marginal operators (both in the charge and spin sectors) of the form $\partial_x\phi_+\partial_x\phi_-$ and $\partial_x\theta_+\partial_x\theta_-$. However, as in Section~\ref{sec:model}, 
these operators can be ignored since their main effect is a renormalization of the scaling dimensions\cite{emery-2000} which here translate in smooth changes of the phase diagrams (without changing their topology) and in the spin gap phases they have essentially no effect.

The first  two lines of Eq.\eqref{eq:Heff}  is the sum of four different LL Hamiltonians for  for the two charge and spin sectors. 
The third line corresponds to different spin backscattering processes, while the fourth and fifth lines represent the singlet and triplet SC couplings and the 
$Q=2(k_1+k_2)=2\pi$ Umklapp processes respectively. 
In addition to the relation between initial value of the luttinger parameters of different sectors and the couplings in the various current-current interaction given in 
Eq.~\eqref{eq:Ks-}, the spin $SU(2)$ invariance dictates that $g_5=g_4+g_3$ and $g_{u5}=g_{u4}+g_{u3}$. 
This will be useful in the discussion of the RG equations and phase diagram.

The Hamiltonian of Eq.\eqref{eq:Heff} has several symmetries. Similarly to the half-filled bonding band case discussed in the preceding sections, 
we find a duality symmetry in the $s-$ spin sector, $(\phi_{s-},\theta_{s-})\rightarrow (\theta_{s-},-\phi_{s-})$,  
under which the Hamiltonian of Eq.\eqref{eq:Heff} retains its form. 
We will denote this symmetry by ${\mathbb Z}^{s-}_2$. However, self-duality holds
only if $g_{s1}=g_{s2}$, $g_{3}=g_4$ and $g_{u3}=g_{u4}$. In addition, the last two lines of  the Hamiltonian of Eq.\eqref{eq:Heff} 
have identical form which indicates that we can define yet another duality symmetry of the same form but this time in the $c-$ charge sector,  
$(\phi_{c-},\theta_{c-})\rightarrow (\theta_{c-},-\phi_{c-})$,  and which will be denoted by ${\mathbb Z}^{c-}_2$. 
Self-duality in this sector requires, in addition, to set $g_5=g_{u5}$.
Finally, the Hamiltonian of Eq.\eqref{eq:Heff} is also even in the fields $\phi_{c,\pm}$, $\theta_{c,\pm}$, $\phi_{s,\pm}$ and $\theta_{s,\pm}$, 
which reflects the invariance under the exchange of the labels  of the Fermi points (or flavors), $1\leftrightarrow 2$,  which is an exact symmetry only for flux $\Phi=\pi$.

In the next section we will look at the different SC and CDW states, each with a unique symmetry properties under the action of the total symmetry group, and construct the order parameters for each state in order to identify the associated quantum phase diagram.

\subsection{Order parameters and phases}
\label{sec:plaquette-order-parameters}

To identify all the phases present in the phase diagram we construct the associated order parameters consistent with the  symmetries of the current problem. In terms of the two flavors $f=1,2$ of the fermions in the bonding band we define for this system two uniform SC order parameters, $\Delta_\pm$, and two PDW order parameters $\tilde \Delta_\pm$ (both with ordering wave vector $Q_{PDW}=\pi$). They are
\begin{align}
\Delta_\pm=& \left(L_{1\up} R_{1\dn}+R_{1\up}L_{1\dn} \right) \pm \left(L_{2\up} R_{2\dn}+R_{2\up}L_{2\dn} \right) \nonumber\\
\tilde \Delta_\pm=& \left(L_{2\up} R_{1\dn}+R_{1\up}L_{2\dn} \right) \pm \left(L_{1\up} R_{2\dn}+R_{2\up}L_{1\dn} \right)
\label{eq:Deltas}
\end{align}
Similarly we also four CDW order parameters, $n_\pm$ and $\tilde n_\pm$,
\begin{align}
n_\pm=& \sum_{\sigma} \Big(L^\dagger_{1\sigma} R_{1\sigma} \pm L^\dagger_{2\sigma} R_{2\sigma}\Big) \nonumber\\
\tilde n_\pm=& \sum_\sigma \Big(L^\dagger_{2\sigma} R_{1\sigma} \pm L^\dagger_{1\sigma} R_{2\sigma}\Big)
\label{eq:ns}
\end{align}
and their adjoint operators.
 
 The relation between these order parameters and the microscopic pair fields and CDW fields is as follows. 
 
 The pair fields defined on site $j$ on each leg $i=1,2$ of the ladder, $\Delta^i_j$, on the rung $j$, $\Delta^{12}_j$, and on each  leg $i=1,2$, $\Delta^i_{j,j+1}$,  are defined by
\begin{align}
\Delta_{i,j}=&c_{i,j,\up} c_{i,\j,\dn}\nonumber \\
\Delta^{12}_{j}=& c_{1,j,\up} c_{2,j,\dn}+c_{2,j,\up} c_{1,j,\dn}\nonumber \\
\Delta^i_{j,j+1}=&c_{i,\up}(j)c_{i,\dn}(j+1)+c_{i,\up}(j+1)c_{i,\dn}(j)
\label{eq:pair-fields-pi-flux}
\end{align}
These observables can be written in terms of the slowly varying chiral Dirac fermions $R_{f,\sigma}$ and $L_{f,\sigma}$ (for the two flavors $f=1,2$) in the symmetrized and anti-symmetrized forms (with respect to the exchange of the labels $1$ and $2$ of the legs of the ladder) 
\begin{subequations}
\begin{align}
	\Delta^1_j+\Delta^2_j &\to \sin \xi \; \Delta_+ +  (-1)^{x/a} \tilde\Delta_+   \\
	\Delta^1_j-\Delta^2_j  &\to - \cos \xi  (-1)^{x/a} \tilde\Delta_- \\
	\Delta^{12}_j & \to \Delta_+ + \sin \xi \; (-1)^{x/a}\; \tilde \Delta_+ \\
	\Delta^1_{j,j+1}+\Delta^2_{j,j+1} & \to 2 \sin \xi \; \sin (qa/4)\; \Delta_- \nonumber\\
	                                                      & - (-1)^{x/a} 2 \im \cos(qa/4)\; \tilde \Delta_-\\
	\Delta^1_{j,j+1}-\Delta^2_{j,j+1} &\to   -  (-1)^{x/a} 2 \im \cos \xi \; \cos(qa/4)\; \tilde \Delta_+
\end{align}
\label{eq:pair-fields-pi-flux-low-energy}
\end{subequations}
where $q=2(k_2-k_1)$ and where we have used the definitions of Eq.\eqref{eq:Deltas}.
We see that the SC order parameters $\Delta_\pm$ and $\tilde \Delta_\pm$ represent two different types of uniform SC states and PDW SC states (both with wave vector 
$Q_{PDW}=\pi$)  respectively. These pairs of SC states differ by their symmetry transformations under flavor exchange. It is worth to note that in the flux $\Phi=\pi$ model 
the PDW order parameters are actually bilinears of fermion operators, {\it c.f.} Eq.\eqref{eq:pair-fields-pi-flux-low-energy}. This is in contrast to what we found in the conventional two-leg ladder in 
section \ref{sec:half-filled}, and to the recent results by Berg {\it et al}\cite{berg-2010} in the Kondo-Heisenberg chain, where the PDW order parameter is microscopically 
quartic in fermion operators. In this sense the PDW states of the flux $\Phi=\pi$ two-leg ladder is closer in spirit to the conventional construction of 
FFLO states,\cite{larkin-1964,fulde-1964}even though the spin $SU(2)$ symmetry is preserved here and explicitly broken in the standard FFLO construction.

Similarly we can relate the site $n_{i,j}$ (with $i=1,2$ the leg index) and rung, $n^{12}_j$ electron charge density operators  
\be
n_{i,j}=\sum_\sigma c_{i,j,\sigma}^\dagger c_{i,j,\sigma}, \qquad
n_j^{12}= \sum_\sigma c_{1,j,\sigma}^\dagger c_{2,j,\sigma}={n_j^{21}}^\dagger
\label{eq:microscopic-densities}
\ee
which, after  symmetrizing and anti-symmetrizing with respect to the exchange of the two legs lead to a set of four order CDW parameters, $n_\pm$ and $\tilde n_\pm$,

The relation between the microscopic charge density operators of Eq.\eqref{eq:microscopic-densities} and the slowly varying chiral Dirac fermions 
$R_{f,\sigma}$ and $L_{f,\sigma}$ (with $f=1,2$) is
\begin{subequations}
\begin{align}
n_{1,j}+n_{2,j}      \to & j^0_1+j^0_2+ \nonumber\\
		                  &\sin \xi \; (-1)^{x/a}\; e^{\im qx/2} n_+ + e^{\im qx/2} \tilde n_+ + \textrm{h.c.}\\
n_{1,j}-n_{2,j}        \to &-\cos \xi (j_1^1-j_2^1)- (\cos \xi \;  e^{\im qx/2} \tilde n_- + \textrm{h.c.}\\
n^{12}_j+n^{21}_j \to &  \sin \xi (j_2^0+j_1^0)\nonumber\\
			         &\!\!\!\! + (-1)^{x/a}\; e^{\im qx/2} n_+ + \sin \xi \; e^{\im qx/2} \tilde n_+ + \textrm{h.c.}\\
n^{12}_j-n^{21}_j  \to &- \cos \xi \;  (-1)^{x/a} \; e^{\im qx/2} n_- - \textrm{h.c.}
\end{align}
\end{subequations}
where we used the definitions of Eq.\eqref{eq:ns}, and the usual definitions of the (normal ordered) currents and densities of the Dirac fermions (again with $f=1,2$)
\begin{align}
j^R_f&=\sum_\sigma R^\dagger_{f,\sigma} R_{f,\sigma}, && j^L_f=\sum_\sigma L^\dagger_{f,\sigma} L_{f,\sigma}\nonumber\\
j^0_f&=j^R_f+j^L_f, && j^1_f=j^R_f-j^L_f
\end{align}
We can also define CDW order parameters on the legs of the ladder. However we will not discuss them since it turns out that they can also be expressed in terms of the same four slowly varying observables  $n_\pm$ and $\tilde n_\pm$ and hence do not bring new information.

From these results we see that in general we find both uniform  SC order parameters and PDW order parameters, which always have  a commensurate ordering wave vector $Q_{PDW}=\pi$. The  CDW order parameters are generally incommensurate and have ordering wave vectors $Q_{CDW}=q/2, \pi\pm q/2$ (or, equivalently $k_2-k_1$, $2k_1$ and $2k_2$).

We will now proceed to write down the bosonized expressions of the SC and CDW order parameters. 
 The bosonized expressions of the SC order parameters are
\begin{subequations}
\begin{align}
	\Delta_+ \propto  e^{-\im\sqrt{\pi}\theta_{c+}} &\big\{ \cos(\sqrt{\pi}\theta_{c-})\cos(\sqrt{\pi}\phi_{s+})\cos(\sqrt{\pi}\phi_{s-}) \nonumber \\
									   &+ \im \sin(\sqrt{\pi}\theta_{c-}) \sin(\sqrt{\pi} \phi_{s+}) \sin(\sqrt{\pi}\phi_{s-})\big\}  
									   \label{eq:Delta+-bosonized-pi-flux}\\
	\Delta_- \propto  e^{-\im\sqrt{\pi}\theta_{c+}} &\big\{ \cos(\sqrt{\pi}\theta_{c-})\sin(\sqrt{\pi}\phi_{s+})\sin(\sqrt{\pi}\phi_{s-})  \nonumber \\
									 &+ \im \sin(\sqrt{\pi}\theta_{c-})\cos(\sqrt{\pi}\phi_{s+})\cos(\sqrt{\pi}\phi_{s-}) \big\} 
									  \label{eq:Delta--bosonized-pi-flux}\\
\tilde\Delta_+ \propto  e^{-\im\sqrt{\pi}\theta_{c+}} & \big\{- \cos(\sqrt{\pi}\phi_{c-})\cos(\sqrt{\pi}\phi_{s+})\cos(\sqrt{\pi}\theta_{s-})  \nonumber \\
									 &+ \im \sin(\sqrt{\pi}\phi_{c-})\sin(\sqrt{\pi}\phi_{s+})\sin(\sqrt{\pi}\theta_{s-}) \big\} 
									  \label{eq:tildeDelta+-bosonized-pi-flux}\\
\tilde\Delta_- \propto  e^{-\im\sqrt{\pi}\theta_{c+}} &\big\{ \cos(\sqrt{\pi}\phi_{c-})\sin(\sqrt{\pi}\phi_{s+})\sin(\sqrt{\pi}\theta_{s-})  \nonumber \\
									 &- \im \sin(\sqrt{\pi}\phi_{c-})\cos(\sqrt{\pi}\phi_{s+})\cos(\sqrt{\pi}\theta_{s-}) \big\}
									  \label{eq:tildeDelta--bosonized-pi-flux}
\end{align}
\end{subequations}
Here, and below, in order to simplify the notation we have dropped the prefactors of these expressions, including the Klein factors, whose effects are taken into account in our results. (A discussion of the role of Klein factors in the identification of phases in ladders is found in ref.[\onlinecite{Marston-2002}].

The  bosonized form of the CDW order parameters $n_\pm$ and $\tilde n_\pm$ are 
 \begin{subequations}
\begin{align}
	n_+ \propto e^{-\im\sqrt{\pi}\phi_{c+}} & \big\{- \cos(\sqrt{\pi}\phi_{c-})\cos(\sqrt{\pi}\phi_{s+})\cos(\sqrt{\pi}\phi_{s-}) \nonumber \\
								&+ \im \sin(\sqrt{\pi}\phi_{c-})\sin(\sqrt{\pi}\phi_{s+})\sin(\sqrt{\pi}\phi_{s-}) \big\} 
								 \label{eq:n+-bosonized-pi-flux}\\
	n_- \propto e^{-\im\sqrt{\pi}\phi_{c+}}  & \big\{ \cos(\sqrt{\pi}\phi_{c-})\sin(\sqrt{\pi}\phi_{s+})\sin(\sqrt{\pi}\phi_{s-}) \nonumber \\
								&- \im \sin(\sqrt{\pi}\phi_{c-})\cos(\sqrt{\pi}\phi_{s+})\cos(\sqrt{\pi}\phi_{s-}) \big\} 
								 \label{eq:n--bosonized-pi-flux}\\
\tilde n_+  \propto e^{-\im\sqrt{\pi}\phi_{c+}}& \big\{ -\cos(\sqrt{\pi}\theta_{c-})\cos(\sqrt{\pi}\phi_{s+})\cos(\sqrt{\pi}\theta_{s-}) \nonumber \\
								&+ \im \sin(\sqrt{\pi}\theta_{c-})\sin(\sqrt{\pi}\phi_{s+})\sin(\sqrt{\pi}\theta_{s-}) \big\} 
								 \label{eq:tilden++-bosonized-pi-flux}\\
\tilde n_{-}  \propto  e^{-\im\sqrt{\pi}\phi_{c+}} &\big\{ \cos(\sqrt{\pi}\theta_{c-})\sin(\sqrt{\pi}\phi_{s+})\sin(\sqrt{\pi}\theta_{s-}) \nonumber \\
	 							&- \im \sin(\sqrt{\pi}\theta_{c-})\cos(\sqrt{\pi}\phi_{s+})\cos(\sqrt{\pi}\theta_{s-}) \big\}
								 \label{eq:tilden--bosonized-pi-flux}
\end{align}
\label{eq:ns-bosonized-pi-flux}
\end{subequations}
where we have also dropped the prefactors and  their dependence on the Klein factors.

The effective field theory of Eq.\eqref{eq:Heff} shows that the spin sector $s+$ couples to the two remaining sectors, the charge sector $c-$ and the spin sector $s-$, only 
through terms that involve the operator $\cos(\sqrt{4\pi}\phi_{s+})$ but not the dual field $\theta_{s+}$.  A consequence of this  feature of the effective Hamiltonian is that 
the Luttinger parameter $K_{s+}$ always decreases under the RG flow, as can be seen by an examination of Eq.\eqref{eq:RG-flow-Ks+}, and flows to a regime in which 
$K_{s+} \to 0$. In this regime the field $\phi_{s+}$ is locked and its fluctuations become massive. Hence there is a gap in the spin sector, the field
$\phi_{s+}$ is pinned, and  $\langle \cos(\sqrt{4\pi}\phi_{s+})\rangle \neq 0$ has a non-vanishing expectation value. 

The RG equations given in Appendix \ref{sec:RG-pi-flux} reveal that for the range of parameters of physical interest all the coupling constants 
(including those in Eq.\eqref{eq:Heff-cminus-sminus}) generically flow to strong coupling. 
Hence, we expect that the operators $\cos(\sqrt{4\pi}\phi_{s-})$, $\cos(\sqrt{4\pi}\theta_{s-})$, $\cos(\sqrt{4\pi}\phi_{c-})$, and $\cos(\sqrt{4\pi}\theta_{c-})$ will acquire an 
expectation value and that the fields become locked to the values 
$\phi_{c-}=n_{\phi_{c-}} \sqrt{\pi}/2$, $\theta_{c-}=n_{\theta_{c-}}\sqrt{\pi}/2$, $\phi_{s-}=n_{\phi_{s-}}\sqrt{\pi}/2$, where $n_{\phi_{c-}}$, $n_{\theta_{c-}}$, $n_{\phi_{s-}}$, 
and $n_{\theta_{s-}}$ are integers that can each be even or odd. Depending of this choice the locked states represent different phases.
In addition, we recall that operators involving dual fields cannot have an expectation value simultaneously as this is forbidden by the commutation relations. This leads us 
to the conclusion that in general we will have different phases depending on which fields are locked and to which values. 
We will label the phases by the locked fields: $(\phi_{c-},\phi_{s-},\phi_{s+})$, 
$(\phi_{c-},\theta_{s-},\phi_{s+})$, $(\theta_{c-},\phi_{s-},\phi_{s+})$, and $(\theta_{c-},\theta_{s-},\phi_{s+})$ respectively. 
Thus, in general we will have a total of eight phases characterized by different order parameters. In all these phases only the charge sector $c+$ remains gapless. 
Additional gapless excitations appear at the continuous quantum phase transitions between these different phases.

From the structure of the effective field theory we see that the $c+$ charge sector decouples and remains critical for all values of the parameters. It is an effective Luttinger 
liquid with Luttinger parameter $K_{c+}$ and velocity $v_{c+}$. This sector has the trivially self-duality of the Luttinger models, which guarantees the existence in the 
phase diagram of a dual CDW state for any SC state,  and vice versa. We will denote this duality symmetry by ${\mathbb Z}^{c+}_2$.

\paragraph{Uniform SC phases:} The bosonized expressions of Eq.\eqref{eq:Delta+-bosonized-pi-flux} and Eq.\eqref{eq:Delta--bosonized-pi-flux} 
for the two uniform SC order parameters, $\Delta_\pm$, 
imply that  these operators may exhibit quasi long range order provided that the $c-$ sector is gapped such that the dual field $\theta_{c-}$ 
is pinned and its vertex operator $\cos(\sqrt{\pi}\theta_{c-})$ has a nonzero expectation value. Thus, the uniform SC  $\Delta_+$ phase 
(even under the exchange of the two legs) occurs whenever 
the fields lock to the classical values $(\theta_{c-},\phi_{s-},\phi_{s+})=(0,0,0)$ or $(\theta_{c-},\phi_{s-},\phi_{s+})=(\pi/2,\pi/2,\pi/2)$. 
Similarly, the  uniform SC $\Delta_-$ phase (odd under the exchange of the two legs) 
occurs whenever the fields lock to the classical values 
$(\theta_{c-},\phi_{s-},\phi_{s+})=(0,\pi/2,\pi/2)$ or $(\theta_{c-},\phi_{s-},\phi_{s+})=(\pi/2,0,0)$. 

\paragraph{PDW phases:} The PDW phase $\tilde \Delta_+$ occurs for $(\phi_{c-},\theta_{s-},\phi_{s+})=(0,0,0)$ and 
$(\phi_{c-},\theta_{s-},\phi_{s+})=(\pi/2,\pi/2,\pi/2)$, while the PDW phase $\tilde \Delta_-$ occurs for 
$(\phi_{c-},\theta_{s-},\phi_{s+})=(0,\pi/2,\pi/2)$ and $(\phi_{c-},\theta_{s-},\phi_{s+})=(\pi/2,0,0)$.
As it should, the order parameters $\Delta_\pm$ and $\tilde \Delta_\pm$, which describe PDW phases which are even and odd under the exchange 
of the two legs respectively, 
exhibit power law correlations due to the contributions form the charge $c+$ sector. Comparing 
the bosonized expressions for $\Delta_\pm$ and $\tilde\Delta_\pm$ it is clear that uniform SC phases and PDW phases  are related by the combined dual 
transformation of the two sectors,
 $\mathbb{Z}^{c-}_2  \times \mathbb{Z}^{s-}_{2}$. 
 In this system PDW phases  cannot occur in the absence of Umklapp process available at flux $\Phi=\pi$, and for this reason 
 are absent for other values of the flux.

\paragraph{CDW phases:} Similarly, the CDW phase $n_+$ has quasi long range order if the field that now lock are 
$(\phi_{c-},\phi_{s-},\phi_{s+})=(0,0,0)$ or $(\phi_{c-},\phi_{s-},\phi_{s+})=(\pi/2,\pi/2,\pi/2)$, the phase 
$n_-$ for $(\phi_{c-},\phi_{s-},\phi_{s+})=(0,\pi/2,\pi/2)$ or $(\phi_{c-},\phi_{s-},\phi_{s+})=(\pi/2,0,0)$, the phase 
$\tilde n_+$ for $(\theta_{c-},\theta_{s-},\phi_{s+})=(0,0,0)$ or $(\theta_{c-},\theta_{s-},\phi_{s+})=(\pi/2,\pi/2,\pi/2)$, and 
$\tilde n_-$ for $(\theta_{c-},\theta_{s-},\phi_{s+})=(0,\pi/2,\pi/2)$ or $(\theta_{c-},\theta_{s-},\phi_{s+})=(\pi/2,0,0)$.

\begin{figure}[hbt]
		\includegraphics[width=0.3 \textwidth]{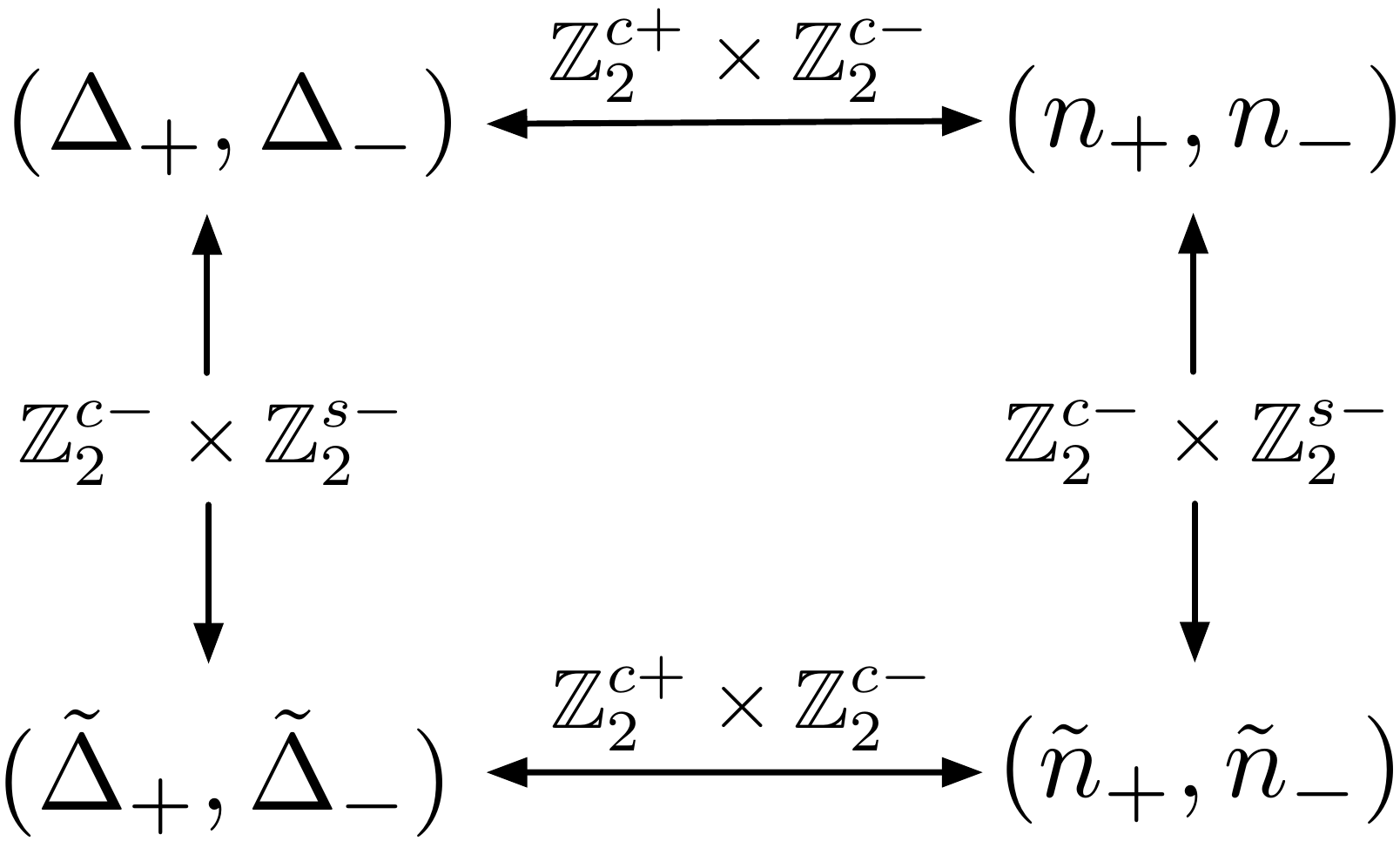}
		\caption{The relation between various uniform and staggered SC and their CDW counterparts present in the phase diagram of the model with the flux 
		$\Phi=\pi$ per plaquette.}
		\label{fig:plaquette-phase-diagram}
	\end{figure}

The diagram of Fig.\ref{fig:plaquette-phase-diagram} illustrates the symmetry relations between various order parameters.

\subsection{Quantum Phase Transitions}

The effective field theory of the ladder with flux $\Phi=\pi$ given in Eq.\eqref{eq:Heff} has many effective parameters and coupling constants. We will not attempt to give a 
detailed description of this theory here. Some important details are given in the Appendices. In particular the RG equations for the effective field theory are given in 
Appendix \ref{sec:RG-pi-flux} and their solution for general couplings is a complex problem. 
From the simpler case of the standard ladder we know that there is always a regime in which the couplings flow to strong values which in this case also corresponds to a 
system with a spin gap. The situation is very similar here. Thus while there are regimes in which some sectors can remain gapless, there is also a generic regime in 
which only one sector, the charge $c+$ sector, remains gapless while all the other ones are massive.

Let us now look at the effective field theory under the assumption that the $s+$ sector is massive (and hence $\phi_{s+}$ is pinned). We will now examine in detail the 
dynamics of the two remaining sectors, the charge sector $c-$ and the spin sector $s-$. In this regime Eq.\eqref{eq:Heff} reduces to the simpler system (ignoring for now 
the decoupled and critical charge sector $c+$)
\begin{align}
\begin{split}
	{\cal H} &=\frac{v_{c-}}{2} \big\{ K_{c-} (\partial_x\theta_{c-})^2 + K^{-1}_{c-} (\partial_x\phi_{c-})^2 \big\}\\
	&+ \frac{v_{s-}}{2} \left\{ K_{s-} (\partial_x\theta_{s-})^2 + K^{-1}_{s-} (\partial_x\phi_{s-})^2 \right\}\\
	&+g^*_{s1} \cos(\sqrt{4\pi}\phi_{s-}) + g^*_{s2} \cos(\sqrt{4\pi}\theta_{s-}) \\
	&+ g^*_5 \cos(\sqrt{4\pi} \phi_{c-}) +g^*_{u5} \cos(\sqrt{4\pi} \theta_{c-}) \\
		&+ \frac{\cos(\sqrt{4\pi}\theta_{c-})}{2(\pi a)^2} \left[ g_{3} \cos( \sqrt{4\pi} \theta_{s-} ) + g_{4} \cos(\sqrt{4\pi} \phi_{s-})  \right]\\
		&+ \frac{\cos(\sqrt{4\pi}\phi_{c-})}{2(\pi a)^2} \left[ g_{u3} \cos( \sqrt{4\pi} \theta_{s-} ) + g_{u4} \cos(\sqrt{4\pi} \phi_{s-})  \right]\end{split}
	\label{eq:Heff-cminus-sminus}
\end{align}
where we absorbed the expectation values of the $s+$ sector in the effective coupling constants $g^*_\alpha = 2g_\alpha\ev{\cos(\sqrt{4\pi}\phi_{s+})}/(2\pi a)^{2}$, where 
$g_\alpha=g_{s1},g_{s2},g_{5},g_{u5}$ respectively.

Let us consider the subspace defined by $g_{s2}=g_3=g_{u3}=0$ in the parameter space.  From the RG equations it can be inferred that once we start with the this initial 
condition, the RG flow will remain on the same hypersurface defined by $g_{s2}=g_3=g_{u3}=0$. 
In this regime luttinger parameters $K_{c\pm}$ and $K_{s\pm}$ follow the RG equations below 
\begin{subequations}
\begin{align}
	&\frac{dK_{c+}}{dl} =0\\
	&\frac{dK_{s+}}{dl} = -\frac{K^2_{s+}}{8\pi^2}(g^2_{s1}+g^2_{5}) \\
	&\frac{dK_{s-}}{dl} = -\frac{K^2_{s-}}{8\pi^2}(g^2_{s1}+g^2_4)\\
	&\frac{dK_{c-}}{dl} = \frac{1}{8\pi^2} (g^2_4+ g^2_5)  - \frac{K^2_{c-}}{8\pi^2}(g^2_{u4}+g^2_{u5})
\end{align}
\end{subequations}
The first equation states that the Luttinger parameter of the decoupled $c+$ sector does not renormalize. 
The second and the third equations state that $K_{s\pm}$ renormalize to small values, $K_{s\pm} \to 0$. 
This means both $s\pm$ sectors are opening up a gap such that the vertex functions of $\phi_{s\pm}$ will acquire expectation values. 
The effective Hamiltonian when the for the regime where $\phi_{s+}$ is pinned is
 \begin{align}
	\begin{split}
 	{\cal H}^{c-}_\text{eff} =& \frac{v_{c-}}{2} \left\{ K^{}_{c-}(\partial_x\theta_{c-})^2+K_{c-}^{-1}(\partial_x\phi_{c-})^2 \right\}\\
		&+\frac{g^{c-}_\theta}{\pi} \cos(\sqrt{4\pi}\theta_{c-})+ \frac{g^{c-}_{\phi}}{\pi}\cos(\sqrt{4\pi}\phi_{c-})
		\label{eq:effective-hamiltonian-c-}
\end{split}
\end{align}
in which $g^{c-}_{\theta}=\frac{{\cal C}_{s+}}{2\pi a}g_{4}$ and $g^{c-}_{\phi}=\frac{{\cal C}_{s+}}{2\pi a}g_{u4}$ where ${\cal C}_{s+} = \ev{\cos(\sqrt{4\pi}\phi_{s+})}$. This 
is the same effective theory of Eq.\eqref{eq:effective-hamiltonian} in  section \ref{sec:PDW-phase-transition} except that it is written for the dual fields in the charge 
$c-$ sector (instead of the spin sector). It predicts existence of a pair of dual phases which between them there is a phase transition in Ising critical class. 
The duality symmetry in the \eqref{eq:effective-hamiltonian-c-} will be denoted by ${\mathbb Z}^{c-}_{2}$. 
It relates the state presented by SC operators $\Delta_{\pm}$ to the states with the same parity presented by $n_\pm$. Similarly $\tilde\Delta_\pm$
 phases and $\tilde n_\pm$ are dual under ${\mathbb Z}^{c+}_{2} \times {\mathbb Z}^{c-}_{2}$. Similar analysis holds in the $s-$ sector. 
 The states with the same parity in $(\Delta_\pm,\tilde n_\pm)$ and $(\tilde
\Delta_\pm,n_\pm)$ are dual under ${\mathbb Z}^{c+}_2\times{\mathbb Z}^{s-}_2$. 

On the other hand, if  we assume that there is relation  between some of the couplings 
(up to restrictions imposed by the $SU(2)$ spin invariance) we arrive to a system that can be solved by refermionization. 
This is discussed in detail in Appendix \ref{sec:refermionization-pi-flux}. Depending on the relations between the coupling constants the system may be in one of the 
phases we discussed above or be qunatum critical. We find two types of quantum criticality. 
One possibility is an ising quantum critical point at which one of the Majorana fermions becomes massless. 
Clearly we have four choices for this. On the other hand we also find a case in which two Majorana fermions become massless. In this case 
the system has a quantum critical regime which can be described as an effective Luttinger model coupled to a massive Thirring model. 
Away from the quantum critical regime this system becomes a theory of four coupled massive Majorana fermions.

\section{Conclusions}
\label{sec:conclusions}

In this paper we investigated the mechanisms of formation of pair-density-wave superconducting order in quasi-one-dimensional systems. 
Although at present time the existence and relevance of the PDW state to the interpretation of experiments in the cuprate superconductors can 
be argued on purely phenomenological grounds, we know that this is not a state that is naturally favored in a weak coupling BCS theory. 
The main motivation of this work is to investigate the mechanisms of formation of PDW order. For this reason it is natural to examine how (and if) 
it it appears on one and quasi-one-dimensional systems.

Here we investigate the occurrence of PDW phases in two models of two-leg ladders. In the first model we reexamined the properties of the spin-gap phase of 
a model of a two-leg ladder in the regime where the microscopic interactions are repulsive and showed that it includes a phase with PDW order.
Here we showed  that within the repulsive regime, a PDW state exists provided that one of the bands, the bonding band for example, is kept at half filling. 
We showed that in this regime the phase diagram of the ladder has, in addition to a conventional Luttinger liquid phase, two superconducting phases: 
a phase with uniform superconducting order (with power law correlations) and a PDW phase, a superconducting state (again with power law correlations) 
but with wave vector $Q_\text{PDW}=\pi$. We also investigated the nature of the quantum phase transition between these two superconducting states and 
showed that it is in the Ising universality class. We discussed in detail the connections that exist between this system and the Kondo-Heisenberg chain. In particular, 
much as in the case of the Kondo-Heisenberg chain, the PDW order parameter in the two-leg ladder is a composite operators of two order parameters of the bonding and 
anti-bonding bands which separately have only short range order. Thus this is a highly non-BCS realization of PDW order.
By extending the analysis to the case other commensurate fillings of the bonding band, we showed that the state with PDW order arises in conjunction with the 
development of a commensurate CDW state. In this sense this result embodies the notion of intertwined orders proposed in Ref.[\onlinecite{berg-2009a}]. 

We also investigated the existence of PDW phases in an extended Hubbard-Heisenberg model on a two leg ladder with flux $\Phi$ per plaquette. We showed that 
commensurate PDW phases appears in this system when the flux $\Phi=\pi$ per plaquette. In contrast to the case of the conventional ladder, this realization of  PDW 
order in the flux $\Phi=\i$ ladder can be expressed as a bilinear of fermion operators. In this sense this realization of the PDW state is closer in spirit to the construction of 
FFLO states although in the problem at hand the spin rotational symmetry is kept unbroken at all levels. PDW order also appears at other values of the flux but only when 
certain commensurability conditions are met, just as it is the case in the conventional two-leg ladder.

There are still several interesting open questions. While the results of this work, and the earlier results of Ref.[\onlinecite{berg-2010}], show how the pair-density-wave 
state arises together with a spin gap in a system with repulsive interactions, the ordering wave vector we find is always commensurate. However there is no reason of 
principle for the PDW ordering wave vector to be commensurate. The root of this phenomenon is the magnetic mechanism of the PDW order which is present in both the 
two-leg ladder and in the Kondo-Heisenberg chain. Indeed in both cases the ordering wave vectors of the PDW and of the spin order (even though it becomes short 
ranged by the development of the  spin gap) are the same. On the other hand, it is not possible to have incommensurate magnetic order (even with power law 
correlations) in one dimension with full $SU(2)$ spin rotational invariance. Indeed it is known from work in frustrated one-dimensional systems that the incommensurate 
magnetic state is preempted in one dimension by a dimerized state with a spin gap. Naturally, one way around this problem is to consider systems with a weak magnetic 
anisotropy. At any rate the construction of a system with incommensurate PDW order is an interesting open problem.

\begin{acknowledgments}
We thank Erez Berg and Steven Kivelson  for very stimulating discussions and a previous collaboration which motivated this work, and G. Roux for point us out Refs. \onlinecite{Roux-2007b} and \onlinecite{Roux-2007}.
This work was supported in part by the National Science Foundation, under grants DMR 0758462 and  DMR-1064319 (EF) at the University of Illinois,
and by the U.S. Department of Energy, Division of Materials Sciences under Award No. 
DE-FG02-07ER46453 through the Frederick
Seitz Materials Research Laboratory of the University of Illinois. 
\end{acknowledgments}

\appendix

\section{RG equations for the flux $\Phi=\pi$ model}
\label{sec:RG-pi-flux}

The RG equations for the model with flux $\Phi=\pi$ per plaquette are
\begin{subequations}
\begin{align}
	&\frac{dK_{c+}}{dl} = 0
	\label{eq:RG-flow-Kc+}\\
	&\frac{dK_{s+}}{dl} = -\frac{K^2_{s+}}{8\pi^2}(g^2_{s1}+g^2_{s2}+g^2_{5} +g^2_{u5})
	\label{eq:RG-flow-Ks+}\\
	&\frac{dK_{c-}}{dl} = \frac{1}{8\pi^2}(g^2_3+g^2_4+g^2_5) - \frac{K^2_{c-}}{8\pi^2}(g^2_{u3}+g^2_{u4}+g^2_{u5})
	\label{eq:RG-flow-Kc-}\\
	&\frac{dK_{s-}}{dl} = -\frac{K^2_{s-}}{8\pi^2}(g^2_{s1}+g^2_4+g^2_{u4}) + \frac{1}{8\pi^2}(g^2_{s2}+g^2_3+g^2_{u3})
	\label{eq:RG-flow-Ks-}\\
	&\frac{dg_{s1}}{dl} = (2-K_{s+}-K_{s-})g_{s1}- \frac{g_4g_5}{2\pi} - \frac{g_{u4}g_{u5}}{2\pi}
	\label{eq:RG-flow-gs1}\\
	&\frac{dg_{s2}}{dl} = (2-K_{s+}-\frac{1}{K_{s-}})g_{s2}- \frac{g_3g_5}{2\pi} - \frac{g_{u3}g_{u5}}{2\pi}
	\label{eq:RG-flow-gs2}\\
	&\frac{dg_{3}}{dl} = (2-\frac{1}{K_{c-}}-\frac{1}{K_{s-}})g_3 - \frac{g_{s2}g_5}{2\pi} 
	\label{eq:RG-flow-g3}\\
	&\frac{dg_{4}}{dl} = (2-\frac{1}{K_{c-}}-K_{s-})g_4 - \frac{g_{s1}g_5}{2\pi}
	\label{eq:RG-flow-g4}\\
	&\frac{dg_{5}}{dl} = (2-\frac{1}{K_{c-}}-K_{s+})g_5 - \frac{g_{s1}g_4}{2\pi} - \frac{g_{s2}g_3}{2\pi}
	\label{eq:RG-flow-g5}\\
	&\frac{dg_{u3}}{dl} = (2-K_{c-}-\frac{1}{K_{s-}})g_{u3} - \frac{g_{s2}g_{u5}}{2\pi} 
	\label{eq:RG-flow-gu3}\\
	&\frac{dg_{u4}}{dl} = (2-K_{c-}-K_{s-})g_{u4} - \frac{g_{s1}g_{u5}}{2\pi}
	\label{eq:RG-flow-gu4}\\
	&\frac{dg_{u5}}{dl} = (2-K_{c-}-K_{s+})g_{u5} - \frac{g_{s1}g_{u4}}{2\pi} - \frac{g_{s2}g_{u3}}{2\pi}
	\label{eq:RG-flow-gu5}
\end{align}
\end{subequations}
with the extra constraint $g_5(0)=g_4(0)+g_3(0)$ and $g_{u5}(0)=g_{u4}(0)+g_{u3}(0)$ to guarantee the spin $SU(2)$ symmetry. 
The above set of RG equations, just as the Hamiltonian itself, is invariant under all the duality symmetries defined in Section \ref{sec:flux} as well as 
under the exchange of the two Fermi points, $1 \leftrightarrow 2$.

\section{Refermionized effective field theory for the two-leg ladder with flux $\Phi=\pi$} 
\label{sec:refermionization-pi-flux}

Here we will assume that the ${s+}$ sector is gapped and the the gapless charge sector $c+$ is decoupled. 
The effective Hamiltonian for the coupled $c-$ and $s-$ sectors is 
given in Eq.\eqref{eq:Heff-cminus-sminus}. Here we 
will discuss the refermionized version  of this effective field theory for some special combinations of parameters.
\begin{align}
	{\cal H} &=\frac{v_{c-}}{2} \big\{ K_{c-} (\partial_x\theta_{c-})^2 + K^{-1}_{c-} (\partial_x\phi_{c-})^2 \big\}\nonumber\\
	&+ \frac{v_{s-}}{2} \left\{ K_{s-} (\partial_x\theta_{s-})^2 + K^{-1}_{s-} (\partial_x\phi_{s-})^2 \right\}\nonumber\\
	+&\frac{{\cal C}_{s+}}{2(\pi a)^2}\Big[ g_{s1} \cos(\sqrt{4\pi}\phi_{s-}) + g_{s2} \cos(\sqrt{4\pi}\theta_{s-}) \nonumber\\
	                                &\qquad\qquad  +g_5 \cos(\sqrt{4\pi} \phi_{c-}) +g_{u5} \cos(\sqrt{4\pi} \theta_{c-}) \Big]\nonumber\\
		                       +& \frac{\cos(\sqrt{4\pi}\theta_{c-})}{2(\pi a)^2} \Big[ g_{3} \cos( \sqrt{4\pi} \theta_{s-} ) + g_{4} \cos(\sqrt{4\pi} \phi_{s-})  \Big]\nonumber\\
		                       &\!\!\!\!\!\!\!\! \!\!\!\!\! + \frac{\cos(\sqrt{4\pi}\phi_{c-})}{2(\pi a)^2}\Big[ g_{u3} \cos( \sqrt{4\pi} \theta_{s-} ) + g_{u4} \cos(\sqrt{4\pi} \phi_{s-})  \Big]
\end{align}
where ${\cal C}_{s+}=\ev{\cos(\sqrt{4\pi}\phi_{s+})}$.
Spin $SU(2)$ invariance dictates $g_3+g_4=g_5$ and $g_{u3}+g_{u4}=g_{u5}$. Let us assume that $g_{3}=g_{u3}=g_{4}=g_{u4}=g/2$ which implies that 
$g_{5}=g_{u5}=g$. Moreover let's assume that $g_{s1}=g_{s2}=g_5=g_{u5}=g$, which implies $K_{s-}(0)=1$. For this regime of parameters , and assuming that the 
velocities $v_{c-}=v_{s-}=v$ are equal, the Hamiltonian simplifies to (with $\alpha=c-,s-$)
\begin{align}
	{\cal H}=&\frac{v}{2} \big\{ K_{c-} (\partial_x\theta_{c-})^2 + K^{-1}_{c-} (\partial_x\phi_{c-})^2 \big\}\nonumber\\
	&+ \frac{v}{2} \big\{ (\partial_x\theta_{s-})^2 + (\partial_x\phi_{s-})^2 \big\}\nonumber\\
	&+ \frac{g}{2(\pi a)^2} \sum_{\alpha} \big\{ \cos(\sqrt{4\pi}\phi_{\alpha})
	+\cos(\sqrt{4\pi}\theta_{\alpha}) \big\} +g {\cal I}_{++}
\end{align}
where the fields ${\cal I}_{\sigma\sigma'}$ are defined as
\begin{align}
	{\cal I}_{\sigma\sigma'} = \frac{1}{(2\pi a)^2}& \left[\cos(\sqrt{4\pi}\phi_{c-})+\sigma\cos(\sqrt{4\pi}\theta_{c-})\right]\nonumber\\
	&\times \left[ \cos( \sqrt{4\pi} \phi_{s-} ) +\sigma' \cos(\sqrt{4\pi} \phi_{s-})\right]
\end{align}

We will now assume that also  $K_{c-}=1$. (Below we will relax this assumption.) We can now define two species of chiral Majorana fermions
\begin{align}
	\chi^\alpha_{R}+\im \xi^\alpha_{R} &=\frac{e^{-\im\pi/4}}{\sqrt{\pi a} } e^{\im\sqrt{4\pi}\phi_{R,\alpha}}\nonumber\\
	 \chi^\alpha_{L}+\im \xi^\alpha_{L} &=\frac{e^{\im\pi/4}}{\sqrt{\pi a} } e^{-\im\sqrt{4\pi}\phi_{L,\alpha}}
\end{align}
where $\alpha=c-,s-$. 

It can be shown that the Majorana mass terms have the bosonized form
\begin{align}
	\im \chi^\alpha_{R} \chi^\alpha_{L} =&\frac{1}{2\pi a} \left[ \cos(\sqrt{4\pi} \phi_{\alpha}) + \cos(\sqrt{4\pi}\theta_{\alpha})\right] \nonumber\\
	\im \xi^\alpha_{R} \xi^\alpha_{L} =&\frac{1}{2\pi a} \left[ \cos(\sqrt{4\pi} \phi_{\alpha})-\cos(\sqrt{4\pi}\theta_{\alpha}) \right]
	\label{eq:majorana-mass}
\end{align}
Using the equations above, the total Hamiltonian for this sector reads as (after setting the velocity $v=1$)
\begin{align}
	{\cal H}=& -\frac{\im}{2} \sum_{\alpha=c-,s-}\left( \xi^{\alpha}_{R}\partial_x \xi^\alpha_{R} 
	-  \xi^{\alpha}_{L}\partial_x \xi^\alpha_{L}\right)\nonumber\\
	&- \frac{\im}{2} \sum_{\alpha=c-,s-} \left( \chi^{\alpha}_{R}\partial_x \chi^\alpha_{R} 
	-  \chi^{\alpha}_{L}\partial_x \chi^\alpha_{L}\right)\nonumber\\
	&+ \im M \sum_{\alpha=c-,s-} \chi^\alpha_{R}\chi^\alpha_{L}  \nonumber\\
	&- g \chi^c_{R}\chi^c_{L} \chi^s_{R}\chi^s_{L}
\end{align}
where the Majorana mass is $M=g{\cal C}_{s+}/(\pi a)$. 
Here, $(\xi^{c-}_R,\xi^{c-}_L)$ and $(\xi^{s-}_R,\xi^{s-}_L)$ are two massless Majorana fields. In this case the system is at a quantum critical point. 

We now note that if $K_{c-} \neq 1$ is allowed, 
the refermionized theory now has a Luttinger-Thirring four fermion (current current) coupling term for the fermions in the $c-$ 
sector of the form $\tilde g R^\dagger_{c-}R_{c-} L^\dagger_{c-}L_{c-}=-4 \tilde g \; \chi^{c-}_R \xi^{c-}_R \chi^{c-}_L\xi^{c-}_L$, 
where $\tilde g$ measures the departure from $K_{c-}=1$. 
This term, which in the conventional Luttinger-Thirring model is marginal, in this case mixes the massless sector with the massive sector. 
However if we were to integrate out the massive sector it will induce a marginal operator in the remaining massless fermions. 
We will see below that the same marginal operator is present automatically if we relax some of the relations between the coupling constants. 
For this reason we will ignore these terms for the time being.

On the other hand, if some of the relations between the coupling constants are lifted (but keeping track of the constraints due to the $SU(2)$ spin symmetry) all four 
Majorana fields become separately massive and the system is in one of the phases described in section \ref{sec:plaquette-order-parameters}. In this language we can 
picture  the system becoming quantum critical by turning one or two Majorana fermions massless. The case with one Majorana fermion becoming massless is the Ising 
quantum criticality that we have already discussed.

Let us now focus on the case in which one pair of Majorana fields remains massless. In this case we can build Dirac Fermions out of the Majorana fermions. This 
transformation will mix the charge and spin fields into a new Dirac field. The right- and the left-moving components are defined as
\begin{align}
\begin{split}
	R_1 = \frac{e^{-\frac{\im\pi}{4}}}{\sqrt{2}} (\chi^{c-}_{R} + \im \chi^{s-}_{R} ), 	\quad
	L_1 = \frac{e^{\frac{\im\pi}{4}}}{\sqrt{2}} (\chi^{c-}_{L} + \im \chi^{s-}_{L} ) \\
	R_2 = \frac{e^{-\frac{\im\pi}{4}}}{\sqrt{2}} (\xi^{c-}_{R} + \im \xi^{s-}_{R} ),  	\quad
	L_2 = \frac{e^{\frac{\im\pi}{4}}}{\sqrt{2}} (\xi^{c-}_{L} + \im \xi^{s-}_{L} )
\end{split}
\end{align}
In terms of the new variables, the Hamiltonian reads as
\begin{align}
	{\cal H} = \sum_{i=1,2} \bar\psi_{i} (-\im \gamma^1\partial_x)\psi_i + M \bar\psi_1 \psi_1 +\frac{g}{4}(\bar\psi_1\gamma^\mu \psi_1)^2 
	\label{eq:Heff-massive-thirring+massless-free}
\end{align}
where $\gamma^0=\sigma_x$, $\gamma^1= - \im \sigma_y$, $\gamma^5=\gamma^0\gamma^1=\sigma_z$ and $\psi^\dagger_i = (R^\dagger_i,L^\dagger_i)$. This 
Hamiltonian consists of a free fermion sector described by free Dirac field $\psi_2$ while the dynamics of the $\psi_1$ is described by massive Thirring model. Therefore 
in this regime of parameters the total symmetry associated with the $s-$ and $c-$ sectors is $U(1)\times U(1)$ which includes a global $U(1)$ symmetry and the chiral 
symmetry of the free sector generated by $\gamma_5$. \newline

For the regime of parameters $g_{s1}=g_{u5}=g_1$ and $g_{s2}=g_{5}=g_2$ while $g_{s1}\neq g_{s2}$, the $\psi_2$ Fermion becomes massive. Thus, the effective 
Hamiltonian of Eq.\eqref{eq:Heff-massive-thirring+massless-free} now has two mass terms of the form
\be
	M \bar\psi_1 \psi_1 + m \bar\psi_2 \psi_2
\ee
with masses given by and $M=\frac{{\cal C}_{s+}}{\pi a} g_+$ and $m=\frac{{\cal C}_{s+}}{\pi a} g_-$ where $g_\pm=(g_{s1}\pm g_{s2})/2$. In this language we have four 
phases depending on the 
relative signs of the masses $M$ and $m$ of the two Dirac fermions.

Under the these assumptions, except for the $SU(2)$ invariance, the $\{g_3,g_{4},g_{u3},g_{u4}\}$ are arbitrary. We re-write the four-fermion interactions as
\be
	{\cal H}_{4F} = f_{11} {\cal I}_{++} + f_{12}{\cal I}_{+-} + f_{21}{\cal I}_{-+} + f_{22}{\cal I}_{--}
\ee
where new coupling constants $f_{\sigma\sigma'}$ is related to the $g$'s as
\begin{align}
	\begin{pmatrix} 
	g_3\\ g_4\\ g_{u3}\\ g_{u4} 
	\end{pmatrix} 
	&= \frac{1}{2}
	\begin{pmatrix}
	 +1& +1& +1& +1 \\ +1 & -1& +1&-1\\+1&+1&-1&-1\\+1&-1&-1&+1
	 \end{pmatrix}  
	 \begin{pmatrix} f_{11}\\ f_{12}\\ f_{21}\\ f_{22} 
	 \end{pmatrix}
	\nonumber\\
	   \begin{pmatrix} 
	   f_{11}\\ f_{12}\\ f_{21}\\ f_{22}
	   \end{pmatrix}
	   &=\frac{1}{2} 
	   \begin{pmatrix} 
	   +1& +1& +1& +1 \\ +1 & -1& +1&-1\\+1&+1&-1&-1\\+1&-1&-1&+1
	   \end{pmatrix}
	   \begin{pmatrix} 
	   g_3\\ g_4\\ g_{u3}\\ g_{u4} 
	   \end{pmatrix} 
\end{align}

Using Eq.\eqref{eq:majorana-mass}, one can write ${\cal I}_{\sigma\sigma'}$ in terms of the Majorana fermions. 
${\cal I}_{++}$, as we saw earlier, is just the current-current interaction of the form $(\bar\psi_1\gamma^\mu\psi_1)^2$. 
By similar argument, ${\cal I}_{--}$ is the same type of interaction but for the $\psi_2$ fields. Together they add up to the following
\be 
	f_{11} {\cal I}_{++} + f_{22}{\cal I}_{--} = f_{11} (\bar\psi_1\gamma^\mu\psi_1)^2 +f_{22}(\bar\psi_2\gamma^\mu\psi_2)^2 
\ee
where  $\bar\psi = \psi^\dagger \gamma_0=(L^\dagger,R^\dagger)$. 
Each operator ${\cal I}_{++}$ and ${\cal I}_{--}$ are invariant under the $U(1)\times U(1)$ symmetry composed of a global $U(1)$ symmetry associated 
with charge conservation and the continuous chiral symmetry of each Dirac fermion. 
The mass terms break the chiral symmetry down to a discrete $\mathbb{Z}_2$ symmetry. 

However, the off-diagonal terms ${\cal I}_{+-}$ and ${\cal I}_{-+}$ involve both $\chi$ and $\xi$ fields. In terms of Majorana fermions they read
\begin{align}
	{\cal I}_{+-} = & \chi^{c-}_R\chi^{c-}_L\xi^{s-}_L\xi^{s-}_R\nonumber\\
	{\cal I}_{-+} =&  \chi^{s-}_R\chi^{s-}_L\xi^{c-}_L\xi^{c-}_R
\end{align}
As it turns out 
these terms violate the conservation of fermion number of the Dirac fermions and, for this reason, are more naturally expressed in terms of Majorana fields.

Let us look at the regime of parameters in which the Dirac fermion number violating couplings are absent and set $f_{12}=f_{21}=0$. 
This happens when $g_3=g_{u4}$ and $g_{4}=g_{u3}$. According to the $SU(2)$ spin rotation invariance condition,  in this regime $g_5=g_{u5}$, 
and therefore the Dirac fermion $\psi_1$  remains massless. Assuming $g_{s1}=g_{s2}$ the Hamiltonian will read as
\begin{align}
	{\cal H} =&  \bar\psi_{1} (-\im \gamma^1\partial_x)\psi_1+ \bar\psi_{2} (-\im \gamma^1\partial_x)\psi_2
	+ M \bar\psi_1 \gamma^0 \psi_1 \nonumber\\
	&+ \frac{G_1}{4}(\bar\psi_1\gamma^\mu \psi_1)^2
	 +\frac{G_2}{4}(\bar\psi_2\gamma^\mu \psi_2)^2 
\end{align}
where $G_1=f_{11} = g_3+g_4$ and $G_2=f_{22} = g_3-g_4$.  
The resulting Hamiltonian splits into the Hamiltonian of massless Thirring model  for $\psi_2$ and a massive Thirring model for the $\psi_1$ with mass 
$M=\frac{{\cal C}_{s+}}{\pi a}(g_3+g_4)$. Therefore what has changed with respect to the case when $g_3=g_4=g_{u3}=g_{u4}$ is that the dynamics of $\psi_2$ field is 
now described by the mass-less Thirring model which shares the same $U(1)\times U(1)$ symmetry with the non-interacting case $G_2=0$. 
Thus, under the (less -restrictive) conditions $g_3=g_{u4}$ and $g_4=g_{u3}$, the symmetry is still $U(1)\times U(1)$. 

Therefore in this case the system decouples into a massless Thirring model and a massive Thirring model each with a separate conserved charge current. The massive 
Thirring model is an integrable system which by bosonization can be mapped onto the sine Gordon field theory,\cite{Coleman-1975} in the regime in which the sine 
Gordon term is relevant.  Hence this sector has a spectrum of massive solitons. On the other hand, the massless Thirring model, which is equivalent to a spinless 
Luttinger model, is a quantum critical system with an exactly marginal operator, parametrized by the coupling constant $G_2$. Hence in this case instead of Ising 
quantum criticality we get Luttinger quantum criticality.

%

\end{document}